\documentclass[journal,10pt,twocolumn,]{IEEEtran}
\usepackage{stfloats}
\ifCLASSINFOpdf
  \usepackage[pdftex]{graphicx}
\else
\fi
\usepackage{amsmath}
\usepackage{caption}
\usepackage{subcaption}
\usepackage{algorithm}
\usepackage{algpseudocode}
\usepackage{svg}

\usepackage{amssymb}
\usepackage{mathtools}
\usepackage{amsthm}
\usepackage{multicol}
\usepackage{adjustbox}

\usepackage{booktabs} 
\usepackage{calc}
\usepackage{lscape}
\usepackage[hidelinks]{hyperref}
\usepackage{array}
\begin{document}
\title{Explainable AI in Speaker Recognition -- Attention Map Visualisation and Evaluation} 

\author{Yanze~Xu,
        Mark D. Plumbley,~\IEEEmembership{Fellow,~IEEE}
        and~Wenwu~Wang,~\IEEEmembership{Fellow,~IEEE}% <-this % stops a space
% \thanks{M. Shell was with the Department
% of Electrical and Computer Engineering, Georgia Institute of Technology, Atlanta,
% GA, 30332 USA e-mail: (see http://www.michaelshell.org/contact.html).}% <-this % stops a space
% \thanks{J. Doe and J. Doe are with Anonymous University.}% <-this % stops a space
% \thanks{Manuscript received April 19, 2005; revised August 26, 2015.}}

\thanks{Yanze Xu, Wenwu Wang are in the Centre for Vision, Speech and Signal Processing, University of Surrey, Guildford, GU2 7XH, UK. Mark D. Plumbley is in Department of Informatics, King's College London, London, WC2R 2LS, UK. Corresponding author: Yanze Xu [yanze.xu@outlook.com]}}

\maketitle

\begin{abstract}

Explaining and understanding the decision-making process of artificial intelligence (AI) systems, particularly those implemented by neural networks, falls within the field of explainable AI (XAI). Analogous to the human attention mechanism, neural networks are assumed to possess their own attention mechanisms that selectively process information during decision-making. This work proposes to study one XAI topic: analysing and visualising the attention mechanisms of neural networks. Our experiments are performed on speaker recognition neural networks that are trained to identify speaker identity from a given utterance.

Previous studies have widely used class activation map (CAM)-based methods to analyse and visualise the attention mechanisms of neural networks. Each of these methods produces an attention map for each network input, highlighting which input regions are selectively processed when the speaker recognition network makes decisions. However, the evaluation of attention maps produced by these methods remains largely underexplored. This work systematically reviews an existing attention map evaluation algorithm, establishing key concepts and identifying its shortcomings. On the basis of this existing evaluation algorithm, a new version is then proposed to address the identified shortcomings, called the Modified Randomised Input Sampling for Explanation - Evaluation algorithm (Modified RISE-eval). Using Modified RISE-eval, we evaluate the attention maps produced by two representative CAM-based methods, GradCAM and LayerCAM, applied to a certain speaker recognition network. The evaluation results demonstrate that GradCAM and LayerCAM each exhibit distinct advantages when applied under different experimental conditions in the speaker recognition task.
\end{abstract}


% \fi
\IEEEpeerreviewmaketitle

\section{Introduction}
% I wish you the best of success.
% \IEEEPARstart{M}{\lowercase{a}}chine Learning (ML) and Deep Learning (DL) methods allow models to learn from data and tackle different tasks (e.g. recognition, generation, recommendation)~\cite{mahesh2020machine}. Nowadays, diverse ML-based models and DL-based models (i.e., neural networks) serve as Artificial Intelligence (AI) systems in our daily lives. Some ML-based models and most neural networks are operating like a black box system, lacking information on how they make decisions. 

\IEEEPARstart{E}xplainable AI (XAI) aims to make the decision-making processes of AI systems, particularly those implemented using neural networks~\cite{he2016deep, lecun2015deep}, transparent and understandable to humans~\cite{gunning2019darpa, xu2019explainable, linardatos2020explainable}. In the deep learning field~\cite{he2016deep, lecun2015deep}, neural networks are typically trained in a supervised manner to learn a nonlinear mapping from inputs (e.g. images or signals) to pre-labelled outputs (e.g. classes or values), with the intermediate outputs of this mapping process serving as task-relevant representations of the inputs~\cite{nasteski2017overview}. Some scholars broadly term all representations as activations, while others strictly reserve this term for representations produced by an activation function alone. This paper adopts the broader usage.

To make neural networks explainable and understandable, we are interested in exploring two high-level XAI questions: \textbf{i)} Since humans possess the ability to organise knowledge and information~\cite{lynn2020humans}, we hypothesise that neural networks have a similar capacity, and ask how a network organises its representations of different inputs. \textbf{ii)} Since human attention is the cognitive ability to selectively process relevant stimuli~\cite{johnston1986selective}, we hypothesise that networks possess an analogous attention mechanism, which we define as the computational ability to selectively process relevant information during decision-making, and ask what information is selectively processed by the network. The first question was addressed in our previous work~\cite{xu2026explainable}; this work focuses exclusively on the second. For this second question, we propose an XAI topic of analysing and visualising the attention mechanism of a well-trained neural network, aiming to reveal what information is selectively attended to during decision-making. Our experiments are conducted within the context of speaker recognition, where networks are typically trained to map any given utterance or its spectrogram to a speaker identity (i.e. classify model input as a speaker class)~\cite{cai2018exploring, nagrani2017voxceleb, chung2020defence}.

\begin{figure}[t]
    \centering
    \includegraphics[width=\linewidth]{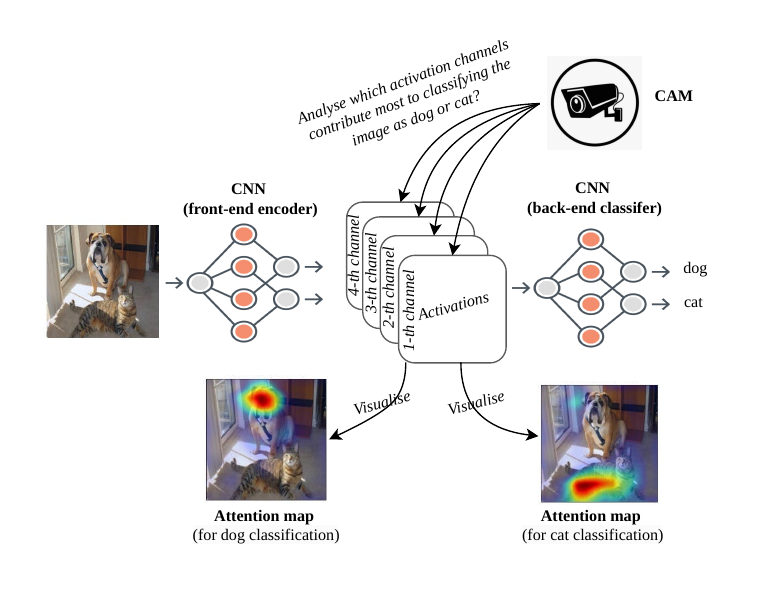}
    \caption{An overview of the Class Activation Map (CAM) method in the image classification task.}
    \label{fig:CAM_quickview}
\end{figure}

Past studies have designed the Class Activation Map (CAM) method~\cite{zhou2016learning} and its variants~\cite{zhou2023master, jiang2021layercam, wang2020score, selvaraju2016grad, jalwana2021cameras} to analyse and visualise the attention mechanism of convolutional neural networks (CNNs) trained for image classification. Specifically, as demonstrated in Fig.~\ref{fig:CAM_quickview}, CAM-based methods first analyse which channels of the intermediate representations (i.e. activations) contribute most to the classification of an input image into a specific class (i.e. dog or cat), and then visualise these activations of different channels as class-specific two-dimensional matrices. Each of these matrices highlights which input regions the network focuses on when making a specific decision (i.e. dog or cat classification), and is referred to as an \textit{attention map}. To this end, we call these CAM-based methods as \textit{attention map visualisation methods}, which can produce attention maps through analysing and visualising the network's attention.

Besides CAM-based methods, a diverse range of attention map visualisation methods has been proposed in the past decade~\cite{lundberg2017unified, ribeiro2016should, 10445989, 10903721, framling2020explainable, petsiuk2018rise, jung2021slrp, vaswani2017attention, Montavon2019}. Since different visualisation methods can produce different attention maps for the same network decision, it is important to evaluate which method better analyses and visualises the network's attention (i.e. produces the better attention map). We collectively refer to such evaluation as \textit{attention map evaluation}. To this end, analysing and visualising a network's attention typically involves first performing attention map visualisation, followed by attention map evaluation.

However, to the best of our knowledge, only a limited number of attention map evaluation methods exist~\cite{jalwana2021cameras, selvaraju2016grad, petsiuk2018rise}, and none has been discussed in detail in the academic literature. To address this gap, this paper provides an in-depth review of an underexplored attention map evaluation algorithm proposed by Petsiuk et al.~\cite{petsiuk2018rise}, which we refer to as the Random Input Sampling for Explanation - Evaluation (RISE-eval). Specifically, RISE-eval uses attention maps as guidance to progressively mask the model input, and evaluates the quality of attention maps by measuring the magnitude of change in model performance, as regions truly selecitvely processed by the network's decision-making will cause a greater performance change when masked.

Our review of RISE-eval not only reveals its core concepts and mechanisms but also uncovers two key evaluation issues: first, under certain conditions, the evaluation results of different attention maps produced by RISE-eval become indistinguishable; second, the evaluation process is confounded by factors unrelated to attention map quality. To this end, we propose the Modified RISE-eval, a new evaluation algorithm built upon RISE-eval, which addresses both of the aforementioned issues through targeted improvements.

Our experiments first analyse and visualise the attention mechanism of a speaker recognition CNN model~\cite{chung2020defence} trained by Chung et al. on the VoxCeleb dataset~\cite{chung2018voxceleb2, nagrani2017voxceleb}. Specifically, two CAM-based attention map visualisation methods are employed: Gradient-weighted Class Activation Mapping (GradCAM)~\cite{selvaraju2016grad} and LayerCAM~\cite{jiang2021layercam}. The resulting attention maps of both methods are visualised for the classification of model inputs (i.e. spectrograms of audio recordings) as target classes (i.e. speaker identities). Subsequently, our proposed Modified RISE-eval is applied to evaluate how well GradCAM and LayerCAM reveal the network's attention, respectively. Our experimental results generally show that GradCAM produces better attention maps when utilising activations from the deepest layer of the examined network, while LayerCAM produces better attention maps when utilising activations from shallower layers.

\section{Background and Related Work} \label{sec:background}
Due to the lack of background discussion about studying the attention mechanism of the neural network, this chapter serves a dual purpose: introducing the necessary background knowledge and reviewing works related to ours.

\subsection{The nature of network attention}
Currently, some AI practitioners assume that only neural networks containing self-attention~\cite{vaswani2017attention, xu2023vision} or similar modules~\cite{santos2016attentive, xue2022vision} possess attention mechanisms. However, our definition of network attention is not limited to networks with specific modules, but rather defines it as the ability of any neural network to selectively process relevant information during the decision-making process, in a manner analogous to human attention.~\cite{johnston1986selective}. Using the concept of human attention to understand neural networks is much like researchers teaching machines to understand images, language, and sounds, all cognitive abilities that were once considered exclusively human.

The essence of network attention lies in the interplay between inputs, gradients, representations (activations), and network parameters. During training, gradients continually update the network parameters; once trained, the network learns to assign greater weight to task-relevant information (i.e. parts of the model inputs or their representations) at each layer, with this information selection compounding across layers~\cite{shwartz2017opening}, such that in the classification layer, parts of the last-layer representation with larger activation values tend to be of greater importance and exert a greater influence on the final output. We therefore propose an approximate hypothesis that the information selectively processed during network decision-making is that which is important and influential to the final decision.
% 不能定稿，最后一句话还有点问题，感觉有点脱离
 % Strictly speaking, however, due to the non-linear nature of neural networks, this hypothesis may not always hold, though it remains broadly reasonable and is not elaborated upon further in this paper.

\subsection{Methods for attention map visusalition}
Attention map visualisation methods analyse and visualise the information selectively processed during network decision-making, presenting the results as a two-dimensional heatmap. We hypothesise that the information selectively processed is equivalent to the information that is important and influential to the final decision. Under this hypothesis, any method that analyses the importance or influence of input regions on model decisions and presents the results as heatmaps is equally considered within the scope of attention map visualisation in this paper.

Some attention map visualisation methods share a similar mechanism: perturbing or masking the input and measuring the influence of the perturbed regions on model decisions, with results presented as heatmaps. Such methods include Local Interpretable Model-agnostic Explanations (LIME)~\cite{ribeiro2016should} and its variants~\cite{10445989}, SHapley Additive exPlanations (SHAP)~\cite{lundberg2017unified}, Randomised Input Sampling for Explanation (RISE)~\cite{petsiuk2018rise}, and Contextual Importance and Utility (CIU)~\cite{10.1007/978-3-030-97546-3_10}. Importantly, these methods do not have to access the internal representations, parameters, or gradients of the neural network, and rely solely on network inputs and outputs.

In contrast, some other attention map visualisation methods directly accesses the internal representations, parameters, and gradients of the neural network. Specifically, Layer-wise Relevance Propagation (LRP)~\cite{Montavon2019} mathematically models the relationship between network inputs and outputs by accessing model parameters, deriving which parts of the input contribute to the output. The self-attention modules~\cite{vaswani2017attention, xu2023vision, ramsauer2020hopfield}, once embedded within a network, can directly map activation values back to the input, under the assumption that positions with larger activation values correspond to the parts of the input that the network processes most prominently. Jacobian-based methods~\cite{1756006.1859912, 10.1007/978-3-319-10590-1_53} compute gradients directly via the chain rule, measuring the influence or contribution of each part of the input or its intermediate representations on the output. CAM-based methods~\cite{selvaraju2016grad, zhou2016learning, jiang2021layercam, zhou2023master, jalwana2021cameras} sit between the self-attention module and Jacobian-based methods: as exemplified by GradCAM~\cite{selvaraju2016grad} and LayerCAM~\cite{jiang2021layercam}, they compute the gradients of intermediate-layer activations with respect to the output to assess each activation's contribution to the final decision, while selectively mapping activations back to the input according to their contribution.

% 总之，由于独立视角的不同，上述各方法所得到的热力图赋予了不同的诠释：网络认为对决策重要的、敏感的、影响较大的、着重关注的（即选择性处理的），或有贡献的。称之为称为feature importance map、sensitivity map、saliency map或attention map。更宏观地研究网络注意力机制的背景下，上述诠释和名称视为等同，可以根据进行选择。

\subsection{Methods for attention map evaluation}

Attention map evaluation methods are considerably fewer in number than visualisation methods, despite the need to evaluate whether attention maps genuinely reflect the network's attention mechanism. One evaluation method~\cite{selvaraju2016grad, chattopadhay2018grad} requires human annotators to pre-label task-relevant regions of the input. Under the assumption that human annotations are correct, this method quantifies how well attention maps localise the human-labelled input regions as a measure of their quality. Another evaluation method, proposed by Petsiuk et al.~\cite{petsiuk2018rise}, takes the following reasoning: since the regions highlighted by the attention map are so-called regions that influence the model's decision, this method verifies this claim by progressively masking these regions from the input and observing whether doing so indeed affects the model decision (i.e. change in model performance). A similar approach is proposed by Jalwana et al.~\cite{jalwana2021cameras}, but employs one-time input masking based on the attention map rather than progressive masking, which reduces computational overhead at the cost of less detailed observation of the influence on the model performance.

Compared to the above works, this paper differs in two respects. First, attention map evaluation in existing works is often treated as a by-product of attention map visualisation methods, receiving little dedicated discussion; this paper, by contrast, places such evaluation within the topic of the network attention mechanism and provides a systematic and dedicated review. Second, while the aforementioned evaluation methods quantify attention map quality, they do not examine whether the evaluation process itself is without issue; this paper identifies inherent limitations in the evaluation method proposed by Petsiuk et al.~\cite{petsiuk2018rise} and proposes an improved version accordingly.

\section{Preliminary Knowledge}
This paper proposes to analyse and visualise network attention by performing attention map visualisation followed by attention map evaluation in the context of a speaker recognition task. This section introduces the background knowledge only relevant to attention map visualisation; in particular, Section~\ref{sec:CAM_knowledge} covers the two CAM-based attention map visualisation methods employed in this paper: Gradient-weighted Class Activation Mapping (GradCAM)~\cite{selvaraju2016grad} and LayerCAM~\cite{jiang2021layercam}.

\subsection{GradCAM and LayerCAM}~\label{sec:CAM_knowledge}
Class Activation Mapping (CAM) method~\cite{zhou2016learning} were originally developed for the image classification task, aiming to analyse and visualise the attention mechanisms of CNN models by measuring the contribution of intermediate activations to the final decision and selectively mapping activations back to the input. GradCAM~\cite{selvaraju2016grad} and LayerCAM~\cite{jiang2021layercam} are two well-known variants of CAM, both originally proposed in the context of image classification, and have since been extended to a broader range of network architectures. CAM-based methods usually refer to their produced heatmaps as saliency maps, following the tradition of early computer vision research~\cite{itti1998model}. In this paper, we adopt the term attention map in keeping with our focus on network attention mechanisms, and provide a detailed discussion of GradCAM and LayerCAM below.

%：既然人类视觉注意力可通过眼动仪刻画其首先关注的区域，那么网络的视觉注意力也应该称为 saliency map

%尽管GradCAM和LayerCAM被广泛使用，现有文献大多只给出公式和简要讨论，鲜有工作深入探讨每个计算步骤背后的物理意义——梯度与激活分别代表什么，两者的组合计算又代表什么，以及最终生成的注意力图应如何被正确诠释。

Let $f$ denote a CNN well-trained for a classification task, such as image classification or speaker recognition, which takes a 2-dimensional input $x$ (e.g.\ an image or audio spectrogram) and produces a predicted class $\hat{y}$. The intermediate convolutional layers of $f$ produce multi-channel 2-dimensional activations as representations of $x$. The 2-dimensional activation of the $k$-th channel is referred to as an activation map $A^k \in \mathbb{R}^{h \times w}$. Here, $h$ and $w$ denote the height and width, $k \in \{1, 2, \dots, K\}$, and $K$ is the total number of channels. The regions with larger activation values within each activation map are considered as the information selectively processed by the network, to generate the activations of the next layer.

These activation maps across different channels are used by GradCAM~\cite{selvaraju2016grad} and LayerCAM~\cite{jiang2021layercam} to determine which channels contribute more to the classification of $x$ into a given class. Specifically, when the given class is $\hat{y}$, GradCAM and LayerCAM compute the gradient of $\hat{y}$ with respect to each spatial location $(i,j)$ of the activation map $A^k$, denoted as $\frac{\partial \hat{y}}{\partial A^k_{i,j}}$. Notably, this gradient is computed with respect to the score of the predicted class $\hat{y}$ only, rather than the scores of any other classes. This ensures that the resulting gradient reflects the contribution or sensitivity of each spatial location of $A^k$ to the score of $\hat{y}$ specifically, rather than any other class.

%a larger graident at position $(i,j)$ indicates that this location contributes more significantly to delivering $\hat{y}$, or equivalently, a smaller perturbation at that location can more strongly influence the predicted score of class $\hat{y}$.

Furthermore, GradCAM next combine activation maps of different channels into a single map, the attention map, as below: 

\begin{equation} 
\widetilde{A^{\hat{y}}}_{\text{grad}} = \text{ReLU}\left(\sum_k \alpha^k_{\hat{y}} A^k\right), \quad \alpha^k_{\hat{y}} = \frac{1}{Z}\sum_i\sum_j \frac{\partial \hat{y}}{\partial A^k_{i,j}}  ~\label{eq:gradcam_fusion}
\end{equation} 

\noindent where $\alpha^k_{\hat{y}}$ is computed by globally average pooling the gradients of $\hat{y}$ with respect to the activation map $A^k$ over all spatial locations, with $Z$ denoting the number of spatial locations. $\alpha^k_c$ serves as the weight of the $k$-th channel, reflecting the sensitivity or importance degree of $k$-th activations to the classification of $x$ as $\hat{y}$. Equation~\eqref{eq:gradcam_fusion} computes a weighted sum of activation maps across all channels, where activation maps of those channels contribute more to the final decision can be preferentially highlighted in the final attention map $\widetilde{A^{\hat{y}}}$. The ReLU function is applied to remove all negative values.

%By doing so, the attention map $L$ takes account not only high-value activation values that useful for learning next-layer activations and useful for outputing the final output.

LayerCAM adopts a different weighted sum of activation maps across all channels in order to genreate the attention map, given as:

% 这里没说正负值

\begin{equation}
\widetilde{A^{\hat{y}}}_{\text{layer}} = \text{ReLU}\left(\sum_k \frac{\partial \hat{y}}{\partial A^k_{i,j}} \odot A^k\right) ~\label{eq:layercam_fusion}
\end{equation}

\noindent where $\odot$ denotes element-wise multiplication. Equation~\eqref{eq:layercam_fusion} directly multiplies the gradient at each spatial location $(i,j)$ with the corresponding activation value element-wise, and sums across all channels to generate the attention map $\widetilde{A^{\hat{y}}}_{\text{layer}}$. By doing so, any spatial location with higher sensitivity or contribution to the final decision (i.e. classification of $x$ as $\hat{y}$) can be preferentially highlighted in the final attention map, rather than requiring the entire channel to have high overall importance as in GradCAM.

%Therefore, the channel-wise weighting of GradCAM preserves the spatial completeness of each important channel's activation map in $L$, while LayerCAM does not.

% activation和graident的共同放缩不说了，直接就直说graidnet放缩吧

Furthermore, the attention maps generated by GradCAM and LayerCAM typically require post-processing: $\widetilde{A^{\hat{y}}}_{\text{grad}}$ and $\widetilde{A^{\hat{y}}}_{\text{layer}}$ are first resized to the resolution of the model input $\mathbf{x}$ (i.e. width $w$ and height $h$) via bilinear interpolation or average pooling, and then normalised to the range $[0, 1]$. Finally, both methods map the highlighted regions of the attention map back to the corresponding spatial regions of the input $x$, under the assumption that the regions of the activations that are contributive or influential for classifying $x$ as $\hat{y}$ are also the same regions in the input $x$ that are important or influential for $f$ classifying $x$ as $\hat{y}$.

\begin{figure*}[h]
    \centering
    \includegraphics[width=0.85\linewidth]{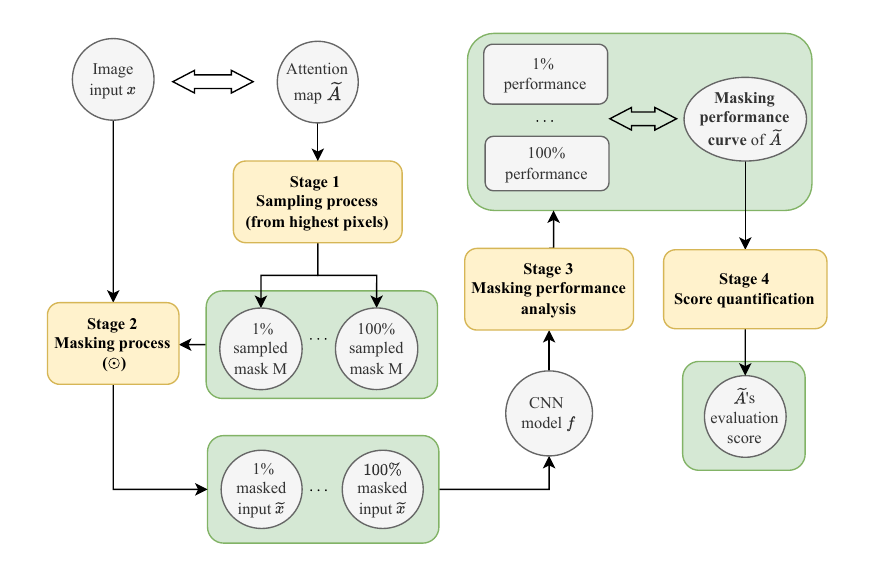}
    \caption{Overview of the Randomised Input Sampling for Explanation - Evaluation (RISE-eval) algorithm~\cite{petsiuk2018rise}, which comprises four stages: sampling attention maps, masking model inputs, analysing model performance during input masking (i.e. referred to as the masking performance analysis), and quantifying scores.} 
    \label{fig:RISE-eval_overview}
\end{figure*}

\section{Methodology}
Analysing and visualising network attention cannot rely solely on attention map visualisation; evaluation of the attention maps is equally necessary to assess their quality and to determine how well they reflect the network's attention mechanism. This section therefore systematically reviews RISE-eval~\cite{petsiuk2018rise}, an underexplored attention map evaluation method proposed by Petsiuk et al., and presents our proposed Modified RISE-eval. Specifically, Sections~\ref{sec:rise_pipeline_concepts}, \ref{sec:rise_eval_implementation}, and \ref{sec:rise_issues} review RISE-eval by outlining its key concepts and evaluation pipeline, providing a detailed implementation, and identifying its evaluation issues, respectively. Building on this review, Section~\ref{sec:modified_rise_method} presents our proposed Modified RISE-eval to address the identified issues.

\subsection{Concepts and pipeline of RISE-eval}\label{sec:rise_pipeline_concepts}

Attention map evaluation aims to assess whether attention map visualisation methods accurately analyse and visualise the network's attention mechanism, whether the resulting attention maps genuinely reflect the network's attention mechanism in other words. However, existing attention map evaluation methods remain scarce. Petsiuk et al.~\cite{petsiuk2018rise} proposed an attention map evaluation method in the context of image classification, which we refer to as Randomised Input Sampling for Explanation - Evaluation (RISE-eval) in this paper. Based on the assumption that attention maps highlight the information selectively processed by the network and considered influential (or important) to the final decision, RISE-eval indeed progressively masks the highlighted regions of model inputs as guided by corresponding attention maps, and quantifies the resulting change in model performance. A greater change in performance indicates higher quality attention maps and the visualisation methods that produce them. However, in Petsiuk et al.'s work, attention map visualisation was the primary focus, and RISE-eval was merely a by-product, receiving little detailed discussion. Subsequently, Li et al.~\cite{li2022reliable} introduced RISE-eval from image classification to speaker recognition, yet provided no methodological discussion. To address this gap, this section provides a thorough review of the concepts and pipeline of RISE-eval. As illustrated in Fig.~\ref{fig:RISE-eval_overview}, the pipeline consists of four stages, each of which is described in turn below.

\textbf{Stage 1\&2 (Sampling and masking processes):} As shown in Fig.~\ref{fig:RISE-eval_overview}, given a 2-dimensional input $x$ and its corresponding attention map $\widetilde{A}$, Stage 1 of RISE-eval first executes a sampling process. Specifically, this sampling process progressively selects the highest-valued pixels from the attention map $\widetilde{A}$ in descending order of pixel values, outputting a binary mask $M$ that indicates which pixels have been selected, with sampled regions set to 1 and the remainder set to 0. $M$ is progressively filled as 1\% to 100\% pixels are progressively sampled from the attention map, generating different versions of $M$ corresponding to different sampling ratios. It is worth noting that in statistics and signal processing, sampling typically refers to random selection. In RISE-eval, however, sampling refers to an ordered selection rather than a random one.

Following the sampling process, RISE-eval then executes the masking process based on the binary mask $M$ and input $x$. Specifically, mask $M$ is applied to model input $x$ based on two strategies: the first directly masks the sampled positions out of the model input, denoted as $\tilde{x} = x \odot M$; the second complementarily masks all regions except the sampled region, denoted as $\tilde{x} = x \odot (1 - M)$, where $\tilde{x}$ is the masked input. Considering the mask $M$ is changing as more and more pixels are sampled from the attention map $\widetilde{A}$, the first strategy progressively deletes the most attentive regions from the original image $x$ until it is fully blank, which is referred to as the \textit{deletion strategy}. The second strategy progressively inserts the most attentive regions into a blank image until it recovers to original input $x$, which is referred to as the \textit{insertion strategy}. Regardless of which strategy is adopted, the masked inputs at masking degrees ranging from 1\% to 100\% are passed to the next stage.

\textbf{Stage 3 (Masking performance analysis process): } The third stage of RISE-eval involves anlaysing how the gradual masking of the model input based on the attention map affects model decisions. Given an input $x$ masked at different degrees, each masked version is independently fed into the network $f$ for classification, and model performance is evaluated at each masking level in terms of classification accuracy or the probability assigned to the ground-truth class. Note that the network $f$ is required to be identical to the one from which the attention map $\widetilde{A}$ is derived, meaning that the same network whose attention mechanism is reflected by $\widetilde{A}$ is used for evaluation. A curve can be drawn to show how model performance changes as an increasing ratio of pixels is used to mask the input, which we refer to as the \textit{masking performance curve} for short. The deletion strategy produces an attention map $\widetilde{A}$ whose masking performance curve monotonically decreases as the mask ratio increases, while the insertion strategy produces a curve that monotonically increases with the mask ratio.

The masking performance curve of $\widetilde{A}$ is comparatively analysed against those of multiple other attention maps visualised for the classification of $x$, in order to reveal the quality of $\widetilde{A}$, which is referred to as the \textit{masking performance analysis}. Specifically, RISE-eval considers a better attention map tends to highlight regions more critical and influential to the model's decision, such that progressively masking those regions in the input leads to a greater performance change. Hence, analysing the descent rate and overall level of masking performance curves of different attention maps allows us to determine which maps cause a greater performance change during the masking process, thereby revealing their respective qualities.
% The masking performances corresponding to masking inputs to different degrees can be drawn as a curve to intuitively observe the change in classification performance during the progressive input masking process, where:
% \begin{itemize}
%     \item The independent variable is the masking degree applied to the image input.
%     \item The dependent variable is the classification performance for each masking degree.
% \end{itemize}
% We refer to this curve as the \textit{masking performance curve}. 

\textbf{Stage 4 (Score quantification process): } The final stage of RISE-eval computes an evaluation score for the attention map based on its masking performance curve. In particular, given the masking performance curve drawn for $\widetilde{A}$, the area under this curve directly serves as the evaluation score of $\widetilde{A}$. The evaluation scores of all attention maps produced by the same visualisation method are then averaged to represent the overall quality of that method.

\subsection{Implementation of RISE-eval} \label{sec:rise_eval_implementation}
Having discussed the key concepts and pipeline of RISE-eval, this section presents its detailed implementation. Notably, since the original code of RISE-eval released by Petsiuk et al.~\cite{petsiuk2018rise} in the context of image classification and the code of RISE-eval reproduced by Li et al.~\cite{li2022reliable} in the context of speaker recognition are both publicly available, we are able to provide a detailed walkthrough of the implementation and identify which algorithmic components can be implemented in different ways.

\begin{algorithm}[t]
    \centering
    \caption{RISE-eval: Sampling and Masking processes}
    \label{alg:naive}
    \begin{algorithmic}[1]
      %-------------- Input & Output -----------------
    \State \textbf{Input:} Input $x_i$ (size $ w \times h$), an attention map $\widetilde{A}_i$ (size $ w \times h$), the number of pixels $n_\mathrm{step}$ to be sampled each iteration, a boolean value $\mathrm{is}_{\mathrm{deletion}}$
    \State \textbf{Output:} The input masked to different degrees $\widetilde{\mathbb{X}}_i$ (size $\frac{w \times h}{n_{\mathrm{step}}} \times w \times h$)%, classification probabilities $P$ (size $\frac{w \times h}{n_\mathrm{step}}$)

    \State \textbf{Initialise: $\widetilde{{X}}_i  = [] $, $M = zeros(w, h)$}
    \While{$sum(M) \neq w \times h$}
        \State
    
        % \State Sample top $n_{\mathrm{step}}$ highest pixels from $\tilde{A_i}$

        % \For{each of the sampled pixels}
        %     \State Set the corresponding position in $M$ to 1 
        % \EndFor
        \State Sample the top $n_{\mathrm{step}}$ pixels with the highest values in $\widetilde{A}_i$, and set their corresponding positions in $M$ to $1$
        % \STATE $M[indexs] = 1$

        \State 
        \If{$\mathrm{is}_\mathrm{deletion} = \text{True}$}
            \State $x_i = x_i \odot M$
        \Else
            \State $x_i = x_i \odot (1 - M)$        
        \EndIf
        \State append $x_i$ to ${\widetilde{X}}_i$

        % \State 
        
        %\State Feed \(\tilde{x}_i\) into model \(f\) in order to obtain the probability on the class $y_i$, and append such probability to \(P\).
        \State 
        \State $\widetilde{A}_i = \widetilde{A}_i 
 \odot M$
    \EndWhile
    \State \textbf{return} ${\widetilde{X}}_i$
    \end{algorithmic}
\end{algorithm}

Algorithm~\ref{alg:naive} presents the pseudocode for the sampling and masking process of RISE-eval. The algorithm takes as input the $i$-th model input $x_i$ and its corresponding attention map $\widetilde{A}_i$, where $i \in \{1, \ldots, n\}$ and $\widetilde{A}_i$ is the attention map produced by a network when classifying $x_i$. In addition, $n_{\mathrm{step}}$ controls the number of pixels sampled and masked in each iteration, and the boolean value $\mathrm{is}_{\mathrm{deletion}}$ specifies whether the deletion or insertion strategy is adopted.

Algorithm~\ref{alg:naive} executes an iterative loop: at each iteration, the top $n_{\mathrm{step}}$ pixels with the highest values in the current attention map are selected to update the masking matrix $M$. Under the deletion strategy (i.e.\ $\mathrm{is}_{\mathrm{deletion}} = \text{True}$), the sampled regions are masked out from $x_i$; under the insertion strategy (i.e.\ $\mathrm{is}_{\mathrm{deletion}} = \text{False}$), all regions except the sampled ones are masked out. All pixels that have been sampled are then removed from the attention map to prevent repeated sampling in subsequent iterations. The loop continues until the model input reaches a masking ratio of $100\%$. The masked inputs at varying masking degrees (i.e. from $0\%$ to $100\%$) are sequentially stored in array $\widetilde{{X}}_i$ and returned as the output of the algorithm.

For the latter two stages of RISE-eval (i.e.\ masking performance analysis and score quantification), Petsiuk et al.\ and Li et al.\ adopt different implementations. In the implementation of Petsiuk et al., the masking performance curve is produced as follows:

\begin{equation}
    c_{\mathrm{Petsiuk}}\!\left(\frac{(j+1) \cdot n_{\mathrm{step}}}{w \times h};\widetilde{A}_i\right) = f\!\left(\widetilde{X}_i[j]\right)[y_i]
\end{equation}

\noindent 

\noindent Here, $c_{\mathrm{Petsiuk}}(\cdot\,;\widetilde{A}_i)$ denotes the masking performance curve of $\widetilde{A}_i$, where each point is defined by feeding each masked version $\widetilde{X}_i[j]$ into the model $f(\cdot)$ and taking the probability of the target class $y_i$ as the $y$-axis, where $j = 0, 1, \ldots, \frac{w \times h}{n_{\mathrm{step}}} - 1$ indexes the masked versions in $\widetilde{\mathbb{X}}_i$. The $x$-axis represents the proportion of masked pixels, where the masking ratio at index $j$ is $\frac{(j+1) \cdot n_{\mathrm{step}}}{w \times h}$, starting from $\frac{n_{\mathrm{step}}}{w \times h}$ and increasing by $\frac{n_{\mathrm{step}}}{w \times h}$ at each step until reaching $1$, at which point the input is completely masked. It is worth noting that the choice of $y_i$ depends on the type of attention map being evaluated. For most attention maps that highlight the regions attended to by $f$ when classifying $x_i$ into its groundtruth class, $y_i$ is set to the groundtruth class of $x_i$. For methods similar to CAM-based methods, where the attention map highlights regions for the classification of $x_i$ as the user-given class, $y_i$ is set to that specified class.

% Here, $j = 0, 1, \ldots, \frac{w \times h}{n_\mathrm{step}}$. 对于某一特定样本 $x_i$，Equation 将 $\widetilde{X}_i$ 中的每个掩蔽版本 $\widetilde{X}_i[j]$ 输入模型 $f(\cdot)$，取其对目标类别 $y_i$ 的概率作为 $y$ 轴。$x$ 轴表示已掩蔽像素占总像素的比例，对应索引 $j$ 的掩蔽比例为 $\frac{j \cdot n_{\mathrm{step}}}{w \times h}$，从 $0$ 开始，每次增加 $\frac{n_{\mathrm{step}}}{w \times h}$，直至 $1$，即输入被完全掩蔽。值得注意的是，$y_i$ 的选择取决于所评估的注意力图的类型。 通常，注意力图高亮网络 $f$ 将 $x_i$ 分类为 groundtruth 类别时所关注的区域， 此时 $y_i$ 为 $x_i$ 的 groundtruth 类别。 对于类 CAM 方法，注意力图高亮的是针对任意指定类别的选择性区域， 此时 $y_i$ 为该指定类别。

For score quantification, the implementation of Petsiuk et al.\ computes the area under the curve (AUC) of the masking performance curve of $\widetilde{A}_i$ as the quality score of this map, denoted as $\mathrm{AUC}(c_{\mathrm{Petsiuk}}(\cdot\,;\widetilde{A}_i))$. The overall quality of the attention maps across all $n$ model inputs is then obtained by averaging their corresponding scores.

Furthermore, the implementation of Li et al. produces the masking performance curve in a different manner, as expressed below:

\begin{multline}
c_{\mathrm{Li}}\!\left(\frac{(j+1) \cdot n_{\mathrm{step}}}{w \times h};\widetilde{\mathbb{A}}\right) \\
    = \frac{1}{n}\sum_{i=1}^{n} \mathbf{1}\!\left[\arg\max f\!\left(\widetilde{X}_i[j]\right) = y_i\right]
\end{multline}

\noindent Unlike Petsiuk et al., whose curve is defined per model input using the probability of the target class, the curve of Li et al.\ (i.e.\ $c_{\mathrm{Li}}(\cdot\,;\widetilde{\mathbb{A}})$) is defined over the entire set of attention maps $\widetilde{\mathbb{A}} = \{\widetilde{A}_i\}_{i=1}^{n}$ corresponding to $n$ model inputs, where the $y$-axis represents the classification accuracy at each masking ratio, computed by averaging the indicator function $\mathbf{1}\!\left[\arg\max f\!\left(\widetilde{X}_i[j]\right) = y_i\right]$ across all model inputs, which evaluates to $1$ if the model correctly classifies the masked input $\widetilde{X}_i[j]$ as $y_i$, and $0$ otherwise. Consequently, $\mathrm{AUC}(c_{\mathrm{Li}}(\cdot\,;\widetilde{\mathbb{A}}))$ directly quantifies the evaluation score of $\widetilde{\mathbb{A}}$, without providing individual scores for individual attention maps.

% Finally, it should be noted that the discussion presented here reflects our own analysis of the RISE-eval method. The original paper does not provide a detailed description of the steps involved in RISE-eval, nor does it mention terms such as the `masking performance curve'. On this basis, to gain further insight, we reviewed the code implementations of RISE-eval proposed by Petsiuk et al.~\cite{petsiuk2018rise} for image classification and implemented by Li \textit{et al.}~\cite{li2022reliable} for speaker recognition. Through this review, we identified that some key differences in their implementations of masking performance analysis:

% \begin{enumerate}
%     \item \textbf{RISE-eval in Image Classification (Petsiuk \textit{et al.}~\cite{petsiuk2018rise}):} This implementation uses the classification probabilities of each individual attention map to plot the masking performance curve.
%     \item \textbf{RISE-eval in Speaker Recognition (Li \textit{et al.}~\cite{li2022reliable}):} In contrast, this implementation uses the classification accuracies of a group of attention maps to plot the masking performance curve.
% \end{enumerate}

% In the following section (Section~\ref{sec:attention_eval_method}), we will discuss this difference in detail while introducing the RISE-eval within the context of speaker recognition.
\begin{figure*}[!t]
    \centering
    \begin{subfigure}[b]{0.45\linewidth}
        \centering
        \includegraphics[width=\linewidth]{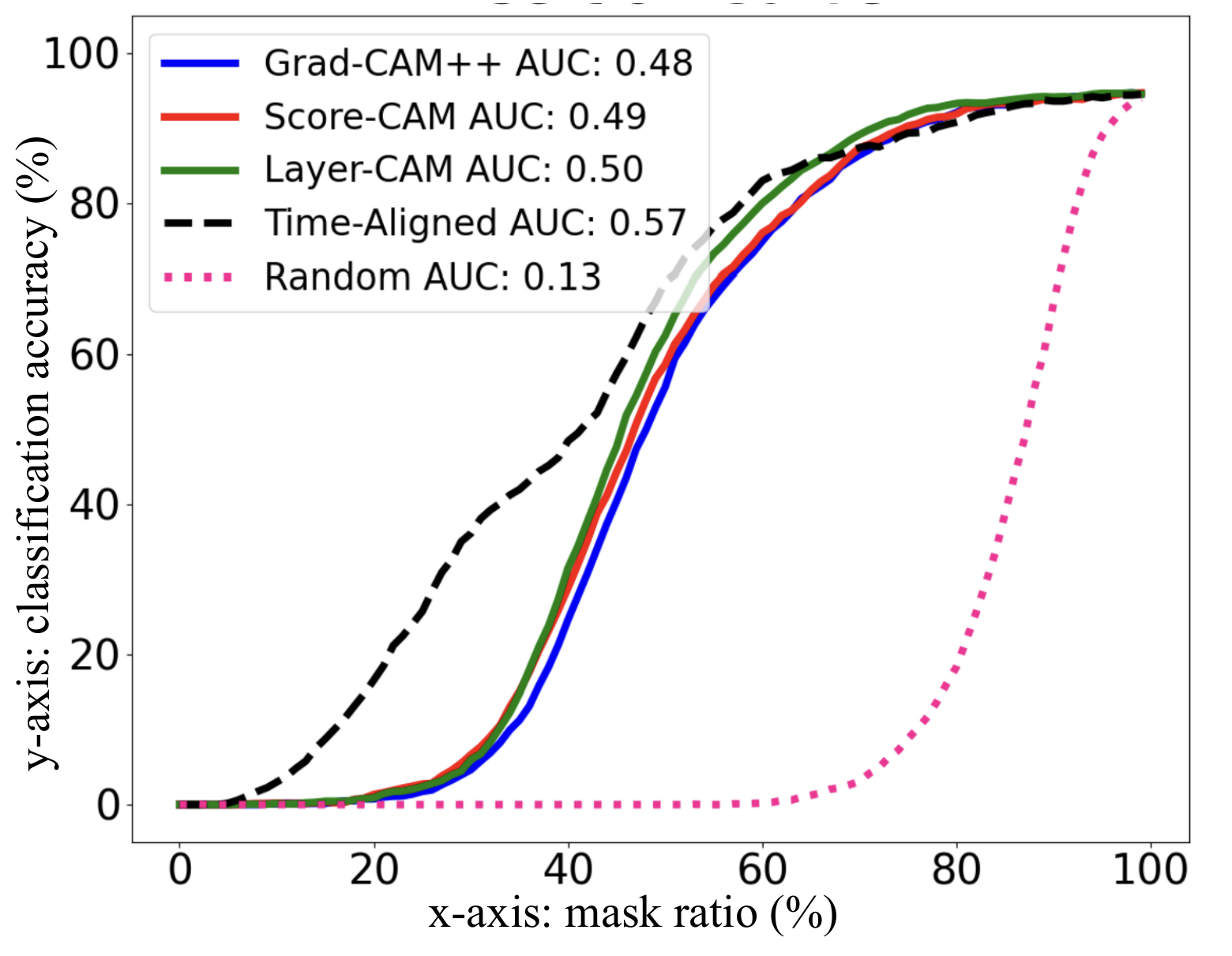}
        \caption{Masking performance curves in insertion strategy}
        \label{fig:li_result_left}
    \end{subfigure}
    \hfill
    \begin{subfigure}[b]{0.48\linewidth}
        \centering
        \includegraphics[width=\linewidth]{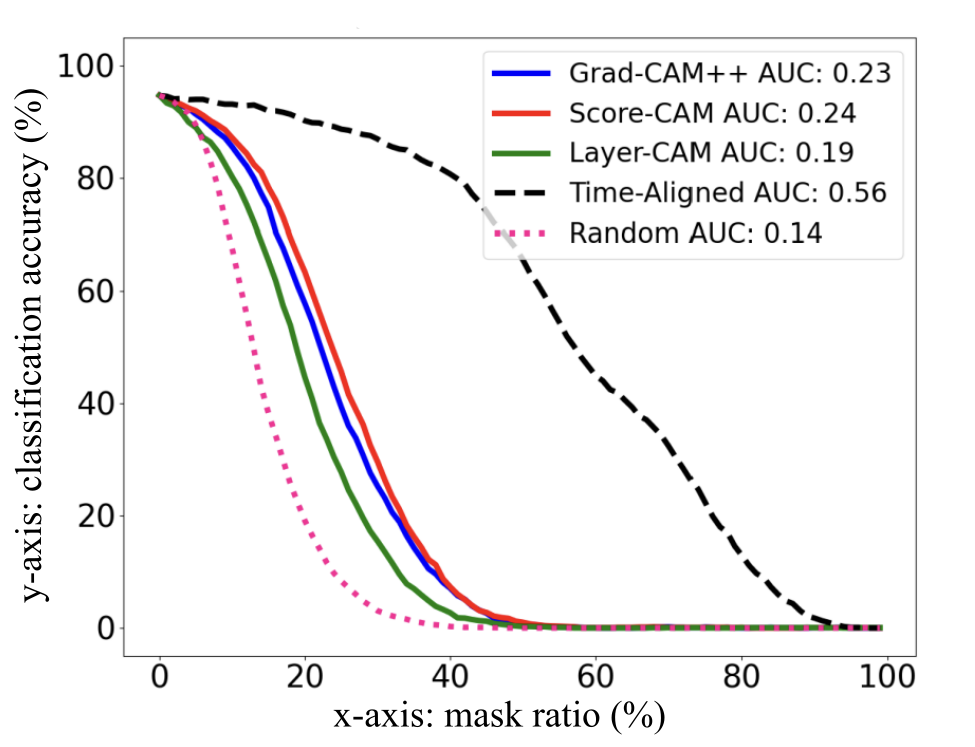}
        \caption{Masking performance curves in deletion strategy}
        \label{fig:li_result_right}
    \end{subfigure}
    \caption{A demonstration of RISE-eval's intermediate evaluation results obtained from Li et al.'s paper~\cite{li2022reliable}.}
    \label{fig:li_result}
\end{figure*}

\subsection{Issues of RISE-eval} ~\label{sec:rise_issues}
In addition to reviewing the pipeline and implementation details of RISE-eval, this section discusses two issues in the evaluation process of RISE-eval through practical examples, and diagnoses which parts of the implementation are responsible for these issues. Fig.~\ref{fig:li_result} shows the masking performance curves produced by Li et al. when applying RISE-eval to evaluate the attention maps produced by three different CAM-based methods in the context of speaker recognition~\cite{li2022reliable}.

Curves in Fig.~\ref{fig:li_result_right} show how model performance (i.e.\ classification accuracy) changes as pixels are progressively deleted from the input (i.e.\ deletion strategy, $\mathrm{is}_{\mathrm{deletion}} = \text{True}$), while curves in Fig.~\ref{fig:li_result_left} show how model performance changes as pixels are progressively revealed in the input (i.e.\ insertion strategy, $\mathrm{is}_{\mathrm{deletion}} = \text{False}$). The blue curve in both subfigures masks inputs according to the attention maps from GradCAM++~\cite{selvaraju2017grad}; the green curve uses maps from LayerCAM~\cite{jiang2021layercam}; and the red curve uses maps from ScoreCAM~\cite{wang2020score}. The remaining two dashed curves in Fig.~\ref{fig:li_result_right} and Fig.~\ref{fig:li_result_left} are not generated from attention maps but are artificially designed anomalous cases, which will not be discussed here.

The first evaluation issue is that the qualities of different attention map visualisation methods are indistinguishable under the insertion strategy. Specifically, the three curves in Fig.~\ref{fig:li_result_left} almost overlap, making it difficult to analyse their level and rate of change, and thus hard to reveal which method produces better attention maps. In contrast, Fig.~\ref{fig:li_result_right} shows clearer separation: the green curve sits at the bottom, followed by the blue and red curves, allowing us to identify that LayerCAM produces better attention maps even before computing the final scores. In summary, this raises the question of whether the insertion strategy of RISE-eval is necessary, as it introduces additional computational overhead without providing effective evaluation results.

\begin{figure}
    \centering
    \includegraphics[width=1\linewidth]{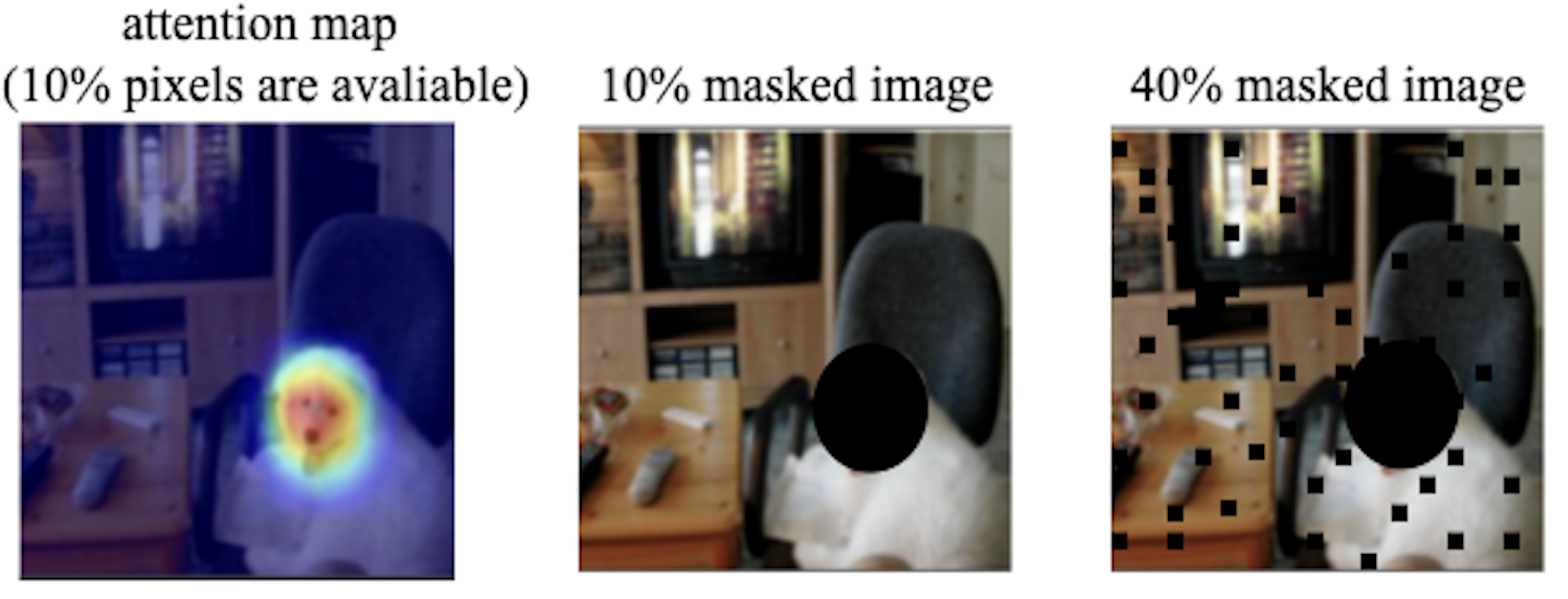}
    \caption{A demonstration for RISE-eval's overmasking of an image input based on its corresponding attention map.}
    \label{fig:overmasking_demo}
\end{figure}

The second evaluation issue is the introduction of irrelevant factors into the evaluation. Specifically, attention maps are inherently sparse~\cite{gardner2024markedly, xu2023visual}: high-value pixels that highlight regions important to the network's decision are scarce, whereas the majority of pixels carry low or zero values, corresponding respectively to regions considered less important or irrelevant to the decision. However, as shown in Fig.~\ref{fig:li_result}, existing algorithms do not stop masking the input until reaching 100\%, which implies the following process: in the early stage, input masking is driven primarily by sampling high-value pixels from the attention map; at some uncertain point in the middle stage, masking begins to be driven by sampling low-value pixels instead; and at some unspecified point in the later stage, pixels with values approaching zero are increasingly sampled, causing the input to become overmasked. Because the numerical differences among these near-zero pixels are negligible, such overmasking becomes approximately equivalent to random masking, meaning the evaluation has drifted away from the attention map and toward a random noise map. Fig.~\ref{fig:overmasking_demo} further illustrates this overmasking effect on the model input. Ultimately, once a sufficiently large number of pixels within the attention map approach zero, the final evaluation score produced by RISE-eval is highly likely to be dominated by interference from the random noise map rather than by the attention map itself, with the root cause being that the input has been overmasked.

% The second  evaluation issue is that RISE-eval的评估过程一致有评估无关的对象干扰。详细的说，先前的研究表明注意力图普遍稀疏，呈长尾分布，这使得大部分工作自然地将注意力图 解读为一种二元选择：高值像素代表网络选择性着重处理的区域，是评估的核心对象； 靠近零的像素由于肉眼几乎无法察觉其被高亮，代表未被网络选择性处理的区域， 是评估无关对象。而RISE-eval无关注意力图被采样的是核心还是无关对象，直接就用来完全掩蔽输入到100\%。产生的连锁问题是：其一，我们无法根据掩蔽性能曲线确定在哪个掩蔽比例之后参与评估的已不再是 核心对象，使得曲线的有效区间难以界定； 其二，无关对象会在打分中会稀释核心对象的贡献——假设核心对象仅占所有像素的 $5\%$，则 $95\%$ 的无关对象将对 AUC 分数的计算占据主导影响，使最终打分反反映的是注意力图未着重高亮区域的质量，而非高亮区域的质量。为此，我们需要在评估中仅针对核心对象进行输入掩蔽和打分，防止利用无关对象过度掩蔽输入。

\subsection{Modified RISE-eval} \label{sec:modified_rise_method}
Referring to the two RISE-eval's issues identified in Section~\ref{sec:rise_issues}, this section presents an improved variant, termed Modified RISE-eval. The main modifications are as follows: for the lack of discrimination in the evaluation results under the insertion strategy, Modified RISE-eval removes this strategy entirely and retains only the deletion strategy; for issues arising from irrelevant factors being introduced into the evaluation, Modified RISE-eval constrains excessive masking to 100\% and introduces a threshold mechanism, whereby pixels in the attention map below the threshold are neither sampled nor used for input masking. We now proceed to discuss the implementation of Modified RISE-eval.

\begin{algorithm}[t]
    \centering
    \caption{Modified RISE-eval: Sampling and masking processes}
    \label{alg: myRISE1}
    \begin{algorithmic}[1]

    \State \textbf{Input:} Input ${x_i}$ (size $w \times h$),
    attention map $\widetilde{A}_i$ (size $w \times h$), importance threshold $t \in [0,1]$, 
    sampling ratio $r^{\text{samp}} \in [0,1]$

    \State \textbf{Output:} Masked input $\widetilde{x}_i$, mask ratio $r^{\text{mask}}$
    \State Initialise $M = \text{zeros}(w, h)$
    \State 
    \State Sample pixels $(u,v)$ satisfying $\widetilde{A}_i[u,v] \geq t$, and set $M[u,v] \leftarrow 1$
    \State 
    \If{$|M| \geq \lfloor w \times h \times r^{\text{samp}} \rfloor$}
        \State $M = \text{zeros}(w, h)$
        \State Sample the top $\lfloor w \times h \times r^{\text{samp}} \rfloor$ pixels with the highest values in $\widetilde{A}_i$, and set their corresponding positions in $M$ to $1$
        \State $r^{\text{mask}} = r^{\text{samp}}$
    \Else
        \State $r^{\text{mask}} = \dfrac{|M|}{w \times h}$
    \EndIf
    \State
    \State $\widetilde{x}_i = {x}_i \odot M$
    \State
    \State \textbf{return} $\widetilde{x}_i$, $r^{\text{mask}}$

    \end{algorithmic}
\end{algorithm}

% \begin{algorithm}[t]
%     \centering
%     \caption{Modified RISE-eval: Sampling and Masking processes}
%     \label{alg: myRISE1}
%     \begin{algorithmic}[1]
%     %-------------- Input & Output -----------------
%     \State \textbf{Input:} Input ${x_i}$ (size $w \times h$),
%     attention map $\widetilde{A}_i$ (size $w \times h$), importance threshold $t \in [0,1]$, 
%     sampling ratio $r^{\text{samp}} \in [0,1]$
%     \State \textbf{Output:} Masked input $\widetilde{x}_i$, mask ratio $r^{\text{mask}}$
%     \State
%     %-------------- Initialisation -----------------
%     \State Initialise $M = \text{zeros}(w, h)$
%     \State
%     %-------------- Mask Generation -----------------
%     \State Sample pixels $(u,v)$ satisfying $\widetilde{A}_i[u,v] \geq t$ and set $M[u,v] \leftarrow 1$
%     \State
%     %-------------- Mask Ratio Computation -----------------
%     \If{$|M| \geq \lfloor w \times h \times r^{\text{samp}} \rfloor$}
%         \State $M$ = \text{zeros}($w$, $h$)
%         \State Sample $\lfloor w \times h \times r^{\text{samp}} \rfloor$ $(u,v)$ from $M$ and update $M$
%         \State $r^{\text{mask}} = r^{\text{samp}}$
%     \Else
%         \State $r^{\text{mask}} = \dfrac{|M|}{w \times h}$
%     \EndIf
%     \State
%     %-------------- Apply Mask -----------------
%     \State $\tilde{x}_i = {x}_i \odot M$
%     \State
%     \State \textbf{return} $\tilde{x}_i$, $r^{\text{mask}}$
%     \end{algorithmic}
% \end{algorithm}

Algorithm~\ref{alg: myRISE1} presents the pseudocode of the sampling and masking procedure in Modified RISE-eval. Given a group of $n$ model inputs $\mathbb{X} = \{x_i\}_{i=1}^{n}$ and their corresponding attention maps $\widetilde{\mathbb{A}} = \{\widetilde{A}_i\}_{i=0}^{n}$, the algorithm takes as input a specific sample $x_i$ and its associated attention map $\widetilde{A}_i$. In addition, the algorithm accepts a threshold parameter $t$, which determines the lower bound for pixel selection; pixels with attention values below this threshold are not sampled for input masking. It also takes a sampling ratio $r^{\mathrm{samp}}$, which specifies the proportion of pixels to be sampled from the attention map for masking the input. The output of the algorithm is the masked input $\tilde{x}_i$ and $r^{\mathrm{mask}}$, which denotes the actual proportion of the input that has been masked.

Algorithm~\ref{alg: myRISE1} firstly marks the binary mask $M$ by sampling all pixels in $\widetilde{A}_i$ whose values are greater than or equal to the threshold $t$; these marked pixels form the candidate pixels. The algorithm then checks whether the number of candidate pixels is sufficient to mask the input to the degree of $r^{\mathrm{samp}}$, i.e., whether $|M| \geq \lfloor w \times h \times r^{\mathrm{samp}} \rfloor$. If sufficient, $\lfloor w \times h \times r^{\mathrm{samp}} \rfloor$ pixels are sampled from $\widetilde{A}_i$ in descending order to update $M$ accordingly, and the actual masking ratio $r^{\mathrm{mask}}$ is set equal to the expected sampling ratio $r^{\mathrm{samp}}$. Otherwise, if $|M| < \lfloor w \times h \times r^{\mathrm{samp}} \rfloor$, all candidate pixels will participate in masking the input, and $r^{\mathrm{mask}}$ is assigned as the ratio between the number of marked pixels in $M$ and the total number of pixels. Lastly, $x_i$ and $M$ are multiplied element-wise to obtain the masked input $\widetilde{x}_i$. Notably, if the number of pixels exceeding the attention threshold is insufficient, the desired sampling ratio cannot be fully achieved, resulting in $r^{\mathrm{mask}} \leq r^{\mathrm{samp}}$.

Furthermore, similar to the implementation of Li et al., Modified RISE-eval generates the masking performance curve by tracking the classification accuracy of a batch of model inputs under varying masking degrees, as shown in Algorithm~\ref{alg: myRISE2}. Specifically, the algorithm takes as input a set of correspondingly paired model inputs, attention maps, and user-specified classes ($X$, $\widetilde{A}$, $Y$); here, $t$ denotes the threshold used to filter pixels for input masking, $R^{\mathrm{samp}}$ is an array of increasing sampling ratios, where each ratio specifies the desired proportion of pixels to be sampled from the attention map for masking the corresponding input, and $f$ is the neural network whose attention mechanism is reflected by the attention map.

\begin{algorithm}[t]
    \centering
    \caption{Modified RISE-eval: Masking performance curve}
    \label{alg: myRISE2}
    \begin{algorithmic}[1]
      %-------------- Input & Output -----------------
    \State \textbf{Input:} Model inputs ${X} = [x_1, \dots, x_n]$, attention maps $\mathbb{\widetilde{A}} = [\widetilde{A}_1, \dots, \widetilde{A}_n]$, classes $Y = [y_1, \dots, y_n]$, threshold $t \in [0,1]$, sampling ratios $R^{\text{samp}}$, neural network $f$
    \State \textbf{Output:} Classification accuracies $P$, averaged mask ratios $R^{\text{mask}}$
    \State \textbf{Initialise:} $P, R^{\text{mask}}$ = []
    \For{each $r^{\text{samp}}$ in $R^{\text{samp}}$}
        \State $\overline{r^{\text{mask}}} = 0$
        \State $p_i = 0$
        \For{each $x_i$ in $X$}
            \State $\tilde{x}_i, r^{\text{mask}} = \textbf{Algorithm\ref{alg: myRISE1}} \; (x_i, y_i, \widetilde{A}_i, t, r^{\text{samp}})$
            \State $\overline{r^{\text{mask}}} = \overline{r^{\text{mask}}} + r^{\text{mask}}$
            \If{$\arg\max f(\tilde{x}_i) = y_i$}
                \State $p_i = p_i + 1$
            \EndIf
        \EndFor
        \State append $(p_i \div n)$ to $P$  
        \State append $(\overline{r^{\text{mask}}} \div n)$ to $R^{\text{mask}}$ 
    \EndFor
    \State \textbf{return} $P$, $R^{\text{mask}}$
    \end{algorithmic}
\end{algorithm}

Algorithm~\ref{alg: myRISE2} constructs the masking performance curve for attention maps $\mathbb{\widetilde{A}}$ through a two-level loop: the outer loop iterates over each sampling ratio $r^{\mathrm{samp}}$ in $R^{\mathrm{samp}}$; for each $r^{\mathrm{samp}}$, the inner loop invokes Algorithm~\ref{alg: myRISE1} to mask each model input in turn, accumulating the mask degree and the classification correctness for each sample. Once the inner loop has finished, the average mask degree and classification accuracy across all model inputs under the current $r^{\mathrm{samp}}$ are appended to the arrays $R^{\mathrm{mask}}$ and $P$, respectively. After the outer loop has iterated over all sampling ratios, $R^{\mathrm{mask}}$ and $P$ together form the horizontal and vertical axes of the masking performance curve, respectively: the former records the average mask ratio achieved by the inputs under each sampling ratio, while the latter records the classification accuracy under the corresponding average masking degree. Notably, using $R^{\mathrm{mask}}$ rather than $R^{\mathrm{samp}}$ as the horizontal axis reveals the effective range of the masking performance curve, namely the range over which the attention map still contains sufficient pixels to support further masking and continued degradation of the model's performance.

Based on the masking performance curve with an effective range, derived for the attention map $\mathbb{\widetilde{A}}$, Modified RISE-eval calculates the final evaluation score of these maps as:

\begin{equation}
\sum_{j=0}^{|P|-2} \frac{P[j] - P[j+1]}{R^{\mathrm{mask}}[j]} ~\label{eq:quantify_score}
\end{equation}

\noindent In Equation~\eqref{eq:quantify_score}, $P$ and $R^{\mathrm{mask}}$ are arrays returned by Algorithm~\ref{alg: myRISE2} for the model inputs $X$ and their corresponding attention maps $\mathbb{\widetilde{A}}$. The numerator, $P[j] - P[j+1]$, captures the drop in the model's classification accuracy as the averaged mask degree increases from $R^{\mathrm{mask}}[j]$ to $R^{\mathrm{mask}}[j+1]$, while the denominator, $R^{\mathrm{mask}}[j]$, rescales this performance change according to the mask degree at which it occurs. Since the elements of $R^{\mathrm{mask}}$ increase monotonically within $[0,1]$, dividing by a smaller, earlier mask degree assigns a larger rescaling weight to the performance change at that step, whereas dividing by a larger, later mask ratio results in a progressively smaller rescaling effect. This design rewards attention maps that can prioritise highlighting the regions and pixels within the inputs that truly influence the model's decisions. Finally, the rescaled performance changes at each step are summed to obtain the final evaluation score for the attention map $\mathbb{\widetilde{A}}$. Notably, a small constant $\epsilon$ should be added to the denominator in implementation, as users may manually set the sampling ratio $r^{\mathrm{samp}}$ to zero at early steps, resulting in $R^{\mathrm{mask}}[0] = 0$ and leading to division by zero if left unaddressed.

\section{Experimental Procedures and Setups} 
\begin{figure*}[t]
    \centering
    \includegraphics[width=0.95\linewidth]{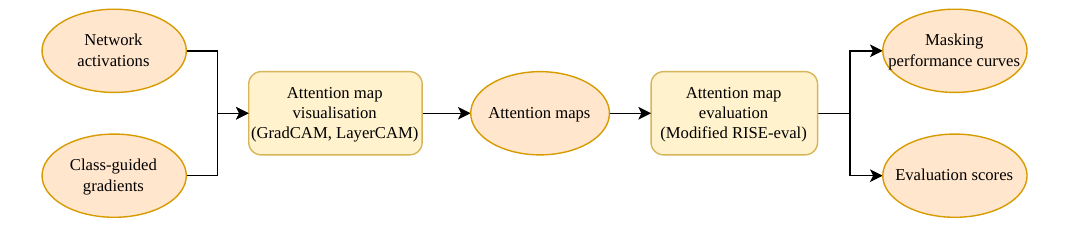}
    \caption{An overview of experimental procedures}
    \label{fig:EXP_procedure}
\end{figure*}

Fig.~\ref{fig:EXP_procedure} generally provides a visual overview of our experimental procedures, while this section presents a detailed description of the setup for each procedure.

\textbf{Preparing network activations and class-guided gradients}: In Fig.~\ref{fig:EXP_procedure}, our experiments begin with obtaining network activations and class-guided gradients for subsequent procedures. Specifically, a ResNet34-based~\cite{he2016deep} speaker recognition network trained by Chung et al.~\cite{chung2018voxceleb2} is downloaded from \href{https://github.com/clovaai/voxceleb_trainer}{https://github.com/clovaai/voxceleb\_trainer}. This network is a CNN trained using prototypical contrastive loss~\cite{snell2017prototypical, chung2020defence} on 200-frame mel spectrograms of 2-second audio clips from the VoxCeleb2 dataset~\cite{chung2018voxceleb2}, achieving an Equal Error Rate (EER) of $2.18\%$ on the VoxCeleb1 test set~\cite{nagrani2017voxceleb} for the task of verifying whether 400-frame mel spectrograms correspond to the correct speaker. As the ResNet34 architecture comprises four ResNet layers, activations are separately extracted from each of these four layers after feeding 200-frame mel spectrograms, computed from 2-second audio clips from the VoxCeleb1 test set~\cite{nagrani2017voxceleb}, into the network.

Furthermore, gradients with respect to the network activations are computed based on the classification  of the above VoxCeleb1's mel spectrograms. Specifically, the gradients are back-propagated only along the pathway corresponding to the ground-truth class probability, reflecting the importance or sensitivity with respect to classifying spectrograms as their ground-truth classes, and such class-specific gradients are computed at the same four ResNet layer positions where activations were extracted.

% Our experiments on attention map visualization, evaluation, and analysis are conducted sequentially. We start by visualizing attention maps from the contrastive-learned CNN model using GradCAM and LayerCAM. Next, we evaluate these attention maps with our modified RISE-eval method, and finally, we analyze maps through both our qualitative and quantitative approaches. This section details the experimental setup used for the attention map visualization and evaluation experiments.

% \textbf{Preparation}: The contrastive-learned CNN model for speaker recognition used in our experiments is sourced directly from the code~\footnote{\url{https://github.com/clovaai/voxceleb_trainer}} provided by JS Chung~\cite{chung2020defence}. This neural network is based on the ResNet34 architecture~\cite{he2016deep} and is trained using prototypical contrastive loss~\cite{chung2020defence, snell2017prototypical}. In more detail, during the training process, the model learns to recognize speaker identities for 200-frame 40 Mel-filter spectrograms extracted from 2-second audio clips in the VoxCeleb 2 training dataset~\cite{chung2018voxceleb2}.

\textbf{Attention map visualisation}: In Fig.~\ref{fig:EXP_procedure}, the prepared gradients and activations are processed by GradCAM and LayerCAM to produce attention maps. As activations and gradients are extracted from four network positions (i.e. four ResNet layers) using two separate methods, this yields a total of eight attention maps for classifying each model input. For GradCAM and LayerCAM, the target class is set to the groundtruth class of each input, since the gradients used in this study correspond to  the groundtruth class. Moreover, the post-processing of the attention maps generated by GradCAM and LayerCAM involves bilinear interpolation for resizing and min-max normalisation for scaling the values.

\textbf{Attention map evaluation}: Given eight groups of attention maps (i.e. GradCAM and LayerCAM applied at four network positions), each showing the network's attention involved in classifying mel spectrograms as their ground-truth classes, the quality of each group is assessed using our proposed Modified RISE-eval. About the setups of Modified RISE-eval, the threshold $t$ is set to $0.2$, such that pixels with attention values below this threshold are excluded from sampling. The sampling ratios $R^{\mathrm{samp}}$ are set to an array including $2\%$, $4\%$, $6\%$, $8\%$, $10\%$, $12\%$, $14\%$, $16\%$, $18\%$, $20\%$, $25\%$, $30\%$, $40\%$, and $50\%$.

\begin{figure*}
    \centering
    \begin{adjustbox}{center}
        \begin{subfigure}{0.75\textwidth}
            \centering
            \includegraphics[width=\textwidth]{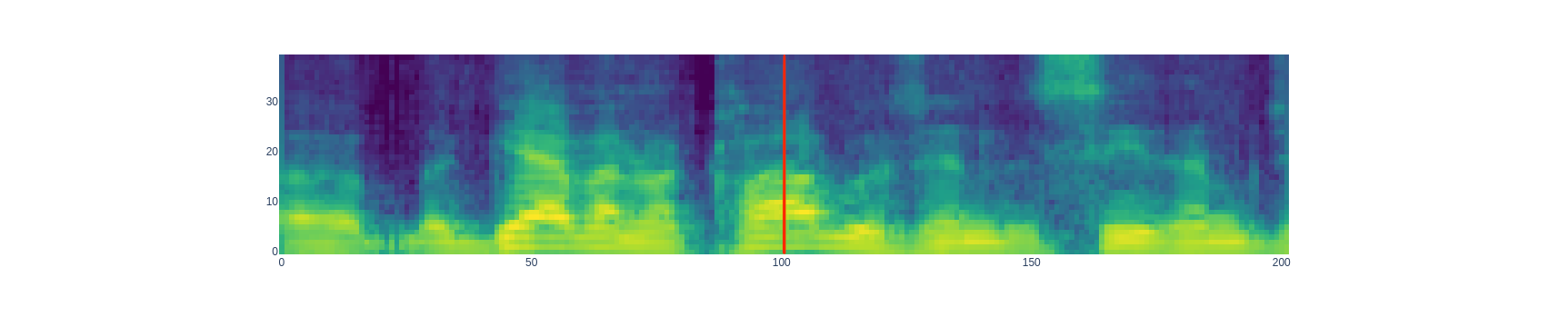}
            \caption{Concatenated mel spectrogram (speaker A on left half | speaker B on right half)}
            \label{fig:combined_a}
        \end{subfigure}
    \end{adjustbox}

    % \vspace{1em}
    \begin{adjustbox}{center}
        \begin{subfigure}[b]{0.5\textwidth}
            \centering
            \makebox[\linewidth][c]{%
                \includegraphics[width=1.5\linewidth]{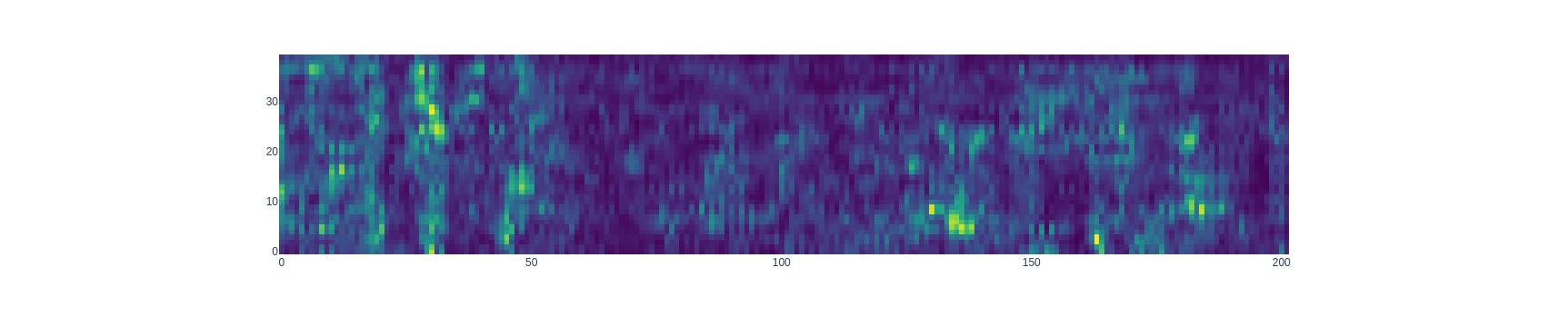}
            }
            \caption{LayerCAM attention map of first ResNet layer (left-half target, 51.68\% energy on the left half)}
            \label{fig:combined_b}
        \end{subfigure}
        \hspace{0.1\textwidth}
        \begin{subfigure}[b]{0.5\textwidth}
            \centering
            \makebox[\linewidth][c]{%
                \includegraphics[width=1.5\linewidth]{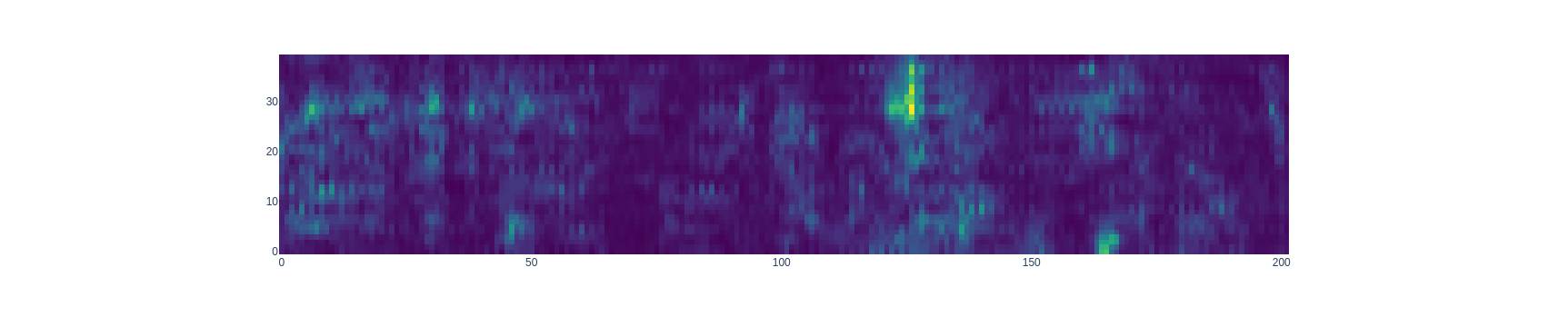}
            }
            \caption{LayerCAM attention map of first ResNet layer (target right half, 46.10\% energy on left half)}
            \label{fig:combined_c}
        \end{subfigure}
    \end{adjustbox}

    % \vspace{1em}
    \begin{adjustbox}{center}
        \begin{subfigure}[b]{0.5\textwidth}
            \centering
            \makebox[\linewidth][c]{%
                \includegraphics[width=1.5\linewidth]{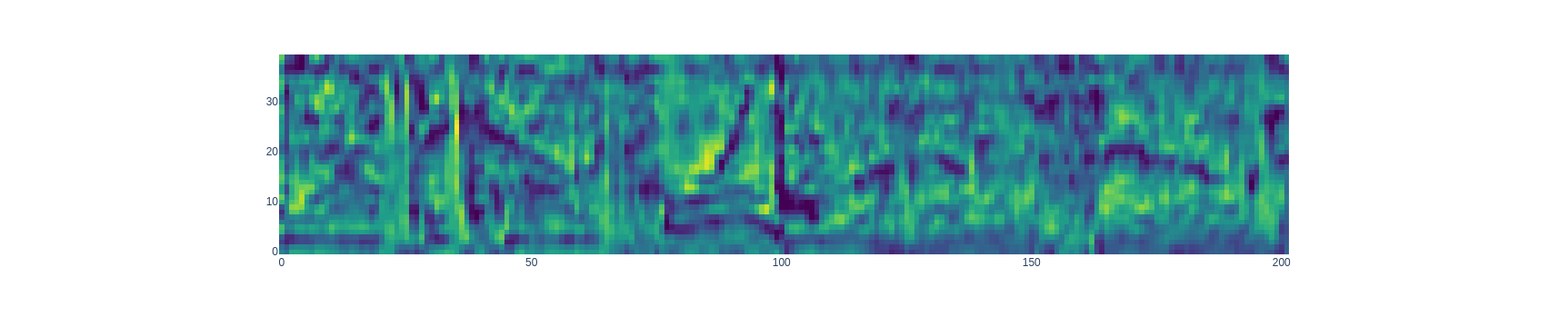}
            }
            \caption{GradCAM attention map of first ResNet layer (target left half, 50.36\% energy on left half)}
            \label{fig:combined_d}
        \end{subfigure}
        \hspace{0.1\textwidth}
        \begin{subfigure}[b]{0.5\textwidth}
            \centering
            \makebox[\linewidth][c]{%
                \includegraphics[width=1.5\linewidth]{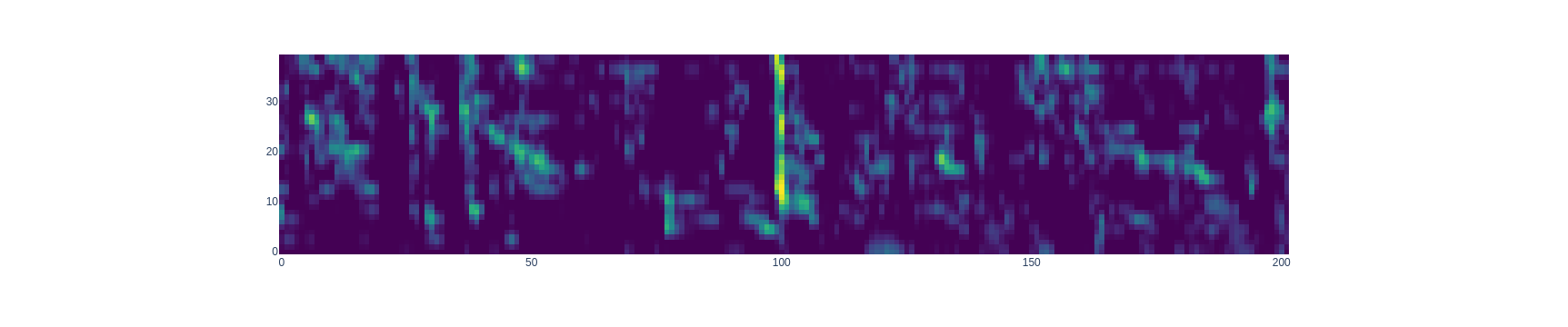}
            }
            \caption{GradCAM's attention map of first ResNet layer (target right half, 53.29\% energy on left half)}
            \label{fig:combined_e}
        \end{subfigure}
    \end{adjustbox}

    % \vspace{1em}
    \begin{adjustbox}{center}
        \begin{subfigure}[b]{0.5\textwidth}
            \centering
            \makebox[\linewidth][c]{%
                \includegraphics[width=1.5\linewidth]{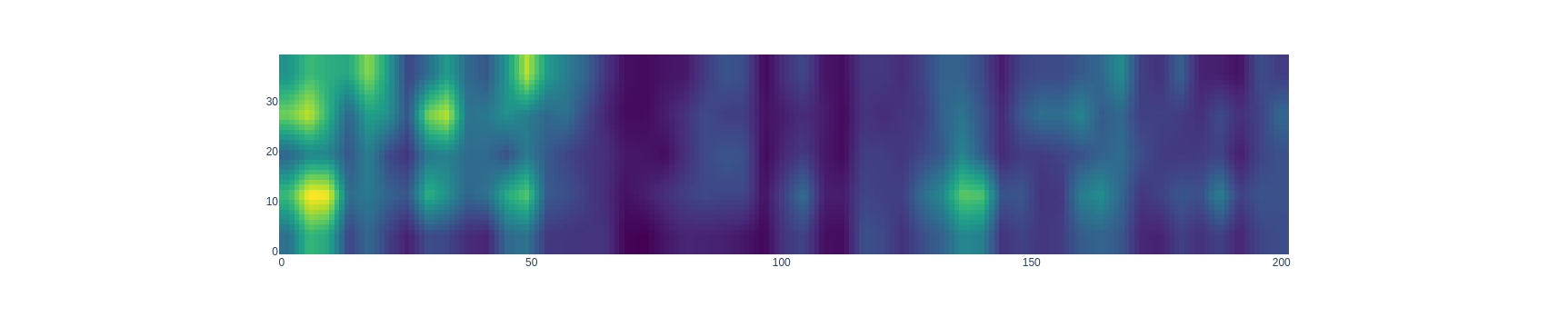}
            }
            \caption{LayerCAM's attention map of last ResNet layer (target left half, 57.66\% energy on left half)}
            \label{fig:combined_f}
        \end{subfigure}
        \hspace{0.1\textwidth}
        \begin{subfigure}[b]{0.5\textwidth}
            \centering
            \makebox[\linewidth][c]{%
                \includegraphics[width=1.5\linewidth]{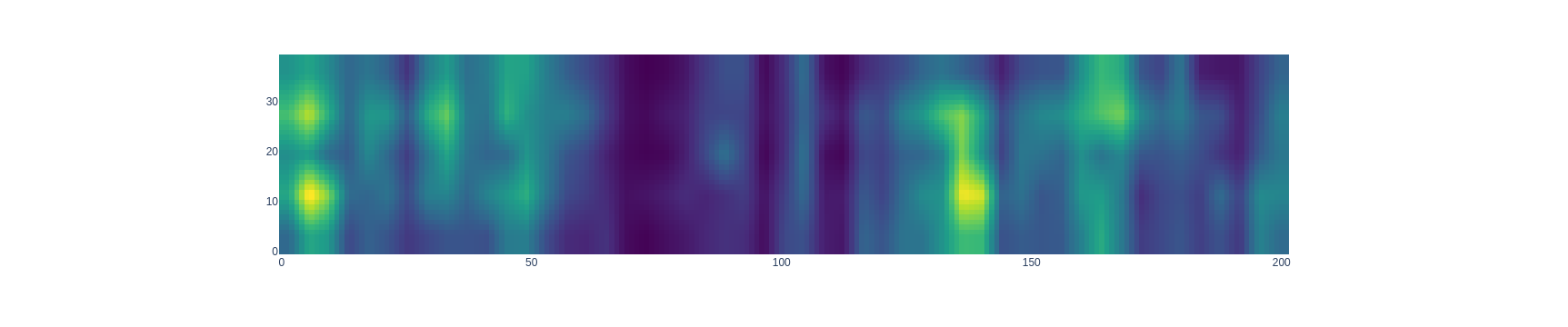}
            }
            \caption{LayerCAM's attention map of last ResNet layer (target right half, 48.18\% energy on left half)}
            \label{fig:combined_g}
        \end{subfigure}
    \end{adjustbox}

    % \vspace{1em}
    \begin{adjustbox}{center}
        \begin{subfigure}[b]{0.5\textwidth}
            \centering
            \makebox[\linewidth][c]{%
                \includegraphics[width=1.5\linewidth]{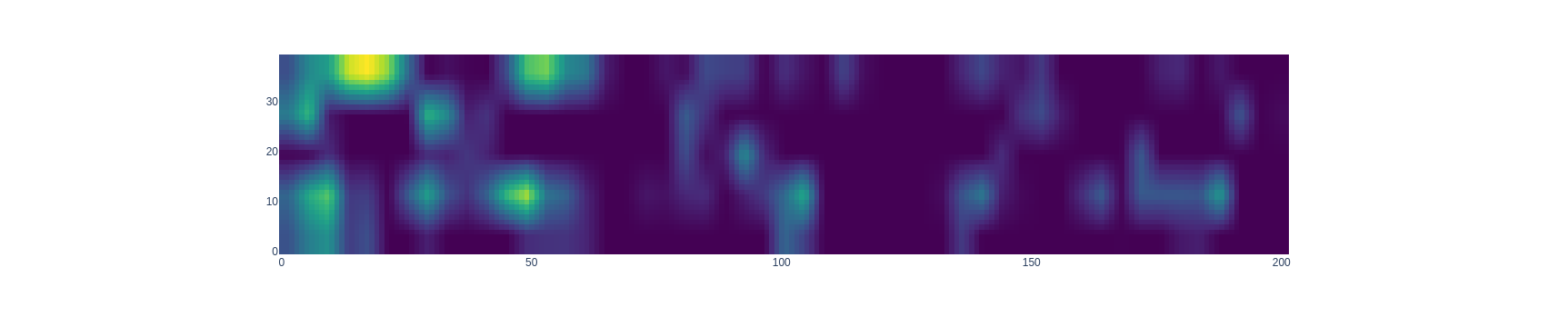}
            }
            \caption{GradCAM's attention map of last ResNet layer (target left half, 76.30\% energy on left half)}
            \label{fig:combined_h}
        \end{subfigure}
        \hspace{0.1\textwidth}
        \begin{subfigure}[b]{0.5\textwidth}
            \centering
            \makebox[\linewidth][c]{%
                \includegraphics[width=1.5\linewidth]{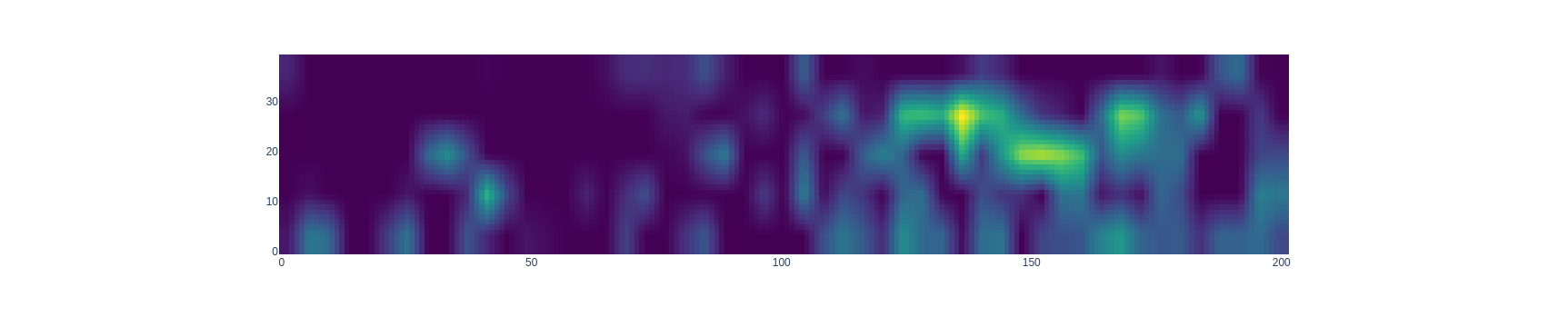}
            }
            \caption{GradCAM's attention map of last ResNet layer (target right half, 19.57\% energy on left half)}
            \label{fig:combined_i}
        \end{subfigure}
    \end{adjustbox}

    \caption{Visualisation of attention maps when our speaker recognition network (i.e. a ResNet34 CNN model) classifies a concatenated mel-spectrogram of a two-speaker utterance as Speaker A (targets the left half of the spectrogram) or Speaker B (targets the right half of the spectrogram). Attention maps are obtained using GradCAM and LayerCAM based on activations from the first and the last ResNet layers that comprise the ResNet34 network.}
    \label{fig:combined_compare}
\end{figure*}

\section{Result analysis}
% This section aims to discuss the results primarily related to the attention map evaluation and analysis. A few attention map visualization results will also be observed, but they will be incorporated into the discussion of the attention map analysis results. For the attention map analysis, subsection xxx demonstrates our approach and results of qualitatively discussing the class-discriminative ability of GradCAM and LayerCAM (applied to the contrastive-learned speaker recognition model) with reference to some specific cases visualized. Subsection xxx demonstrates our approach and results of quantitatively discussing the numerical characteristics of GradCAM and LayerCAM concerning counting distributions for pixels of a large number of attention maps. For the attention map evaluation, subsection xxx demonstrates the results of using the Modified RISE-eval to evaluate GradCAM and LayerCAM.

This section presents our analysis and visualisation of the attention of a speaker recognition network by applying two attention map visualisation methods (i.e. GradCAM and LayerCAM), with attention map evaluation (i.e. Modified RISE-eval) carried out to assess such analysis and visualisation. In particular, Section~\ref{sec:qualitative_analysis} first presents some visual exemplars of the attention maps produced by GradCAM and LayerCAM in the context of the speaker recognition task. Section~\ref{sec:evaluation_analysis} then reports the intermediate evaluation results from Modified RISE-eval for GradCAM and LayerCAM, namely the masking performance curves. Finally, Section~\ref{sec:evaluation_scores} presents the final evaluation scores from Modified RISE-eval, summarising how well GradCAM and LayerCAM analyse and visualise the attention mechanism of the speaker recognition network.

\subsection{Visual demonstration of GradCAM and LayerCAM}~\label{sec:qualitative_analysis}

% In Section xxx, we outline the approach of our qualitative analysis method, building upon the method proposed by Li et al., which considers five possible scenarios to provide a more detailed discussion of the efficiency of the class-discriminative ability of individual attention maps. Additionally, our approach is characterized by comparing the transfer of attentive pixels among diverse attention maps while the model classifies the same sample into different categories, with the goal of more effectively confirming the existence of class-discriminative ability. In this subsection, we use specific examples to better understand our proposed method. We also present the results of analyzing the class-discriminative ability of GradCAM and LayerCAM, of which two visualization methods we applied to the contrastive-learned CNN model.

GradCAM and LayerCAM are two popular CAM-based methods used to analyse and visualise the attention mechanism of our speaker recognition network (i.e. a ResNet34-based CNN). CAM-based methods are characterised by highlighting which parts of the input a CNN selectively processed, or considered important, sensitive, or influential, for classifying the input as a user-given class.

In this section, we follow the approach of Li et al.~\cite{li2022reliable} to demonstrate the effectiveness of applying GradCAM and LayerCAM to our exmained speaker recognition neural network. Specifically, we concatenate mel spectrograms from two random speakers into a single 200-frame spectrogram, where the first 100 frames belong to one speaker (i.e. Speaker A) and the remaining 100 frames belong to the other (i.e. Speaker B), as shown in Fig.~\ref{fig:combined_a}. GradCAM and LayerCAM are then applied to separately highlight the important input regions involved in classifying this concatenated spectrogram as the first and the second speaker, respectively. It is assumed that, if GradCAM and LayerCAM function as intended, their attention maps could selectively highlight the spectral region corresponding to the target speaker when classifying the concatenated spectrogram as that target speaker, such that Speaker A's region is highlighted for a Speaker A classification, and Speaker B's region is highlighted for a Speaker B classification.

Figs.~\ref{fig:combined_b} and \ref{fig:combined_d} show corresponding GradCAM and LayerCAM maps for classifying the concatenated spectrogram as belonging to Speaker A, using activations extracted from the first ResNet layer. Figs.~\ref{fig:combined_f} and \ref{fig:combined_h} show corresponding GradCAM and LayerCAM maps targeting the same speaker, but using activations extracted from the last ResNet layer. Observing Figs.~\ref{fig:combined_b} and \ref{fig:combined_d}, it is difficult to determine whether the first-layer attention maps of GradCAM and LayerCAM preferentially highlight the left or right half of the spectrogram. The proportions of energy located in the left half of Figs.~\ref{fig:combined_b} and \ref{fig:combined_d} are 51.68\% and 50.36\%, respectively, only marginally exceeding 50\%. In contrast, as shown in Figs.~\ref{fig:combined_f} and \ref{fig:combined_h}, the last-layer attention maps of GradCAM and LayerCAM clearly emphasise the left half of the spectrogram corresponding to Speaker A. The left-half energy ratio of Fig.~\ref{fig:combined_f} reaches 57.66\%, and that of Fig.~\ref{fig:combined_h} reaches a slightly higher value of 76.30\%, suggesting that the last-layer attention map of GradCAM more accurately highlights the target speaker's region than that of LayerCAM.

Furthermore, Figs.~\ref{fig:combined_c}, \ref{fig:combined_e}, \ref{fig:combined_g}, \ref{fig:combined_i} respectively share the same experimental setup as Figs.~\ref{fig:combined_b}, \ref{fig:combined_d}, \ref{fig:combined_f}, \ref{fig:combined_h}, with the only difference being that they target Speaker B instead. Observing Figs.~\ref{fig:combined_c} and \ref{fig:combined_e}, it is difficult for us to determine whether the first-layer attention map of GradCAM and the first-layer attention map of LayerCAM preferentially highlight the left or right half of the spectrogram. The proportions of energy located in the left half of Figs.~\ref{fig:combined_c} and \ref{fig:combined_e} are 46.10\% and 53.29\%, respectively, both hovering around 50\%. In contrast, as shown in Figs.~\ref{fig:combined_g} and \ref{fig:combined_i}, the first-layer attention map of LayerCAM fails to clearly distinguish between the left and right halves of the spectrogram, whereas the last-layer attention map of GradCAM highlights the right half far more prominently. The left-half energy ratio of Fig.~\ref{fig:combined_g} reaches 48.18\%, and that of Fig.~\ref{fig:combined_i} reaches a notably lower value of 19.57\%, suggesting that the last-layer attention map of GradCAM more accurately highlights the target speaker's region than the first-layer attention map of LayerCAM.

In summary, a case is discussed about the ability of GradCAM and LayerCAM's attention maps to discriminate and localise the user-specified speaker's region within model inputs, an ability referred to as class-discriminative ability~\cite{selvaraju2016grad}. The attention maps of both GradCAM and LayerCAM show inconsistent class-discriminative ability, sometimes successfully preferring to highlight the target speaker's region and sometimes failing to do so. However, it is worth noting that the last-layer attention map of GradCAM exhibits strong class-discriminative ability, accurately higlighting the target speaker's region and shifting this focus accordingly when the target speaker is switched. Notably, there is a textural pattern among these attention maps: first-layer attention maps tend to capture more fine-grained detail, with their high-value regions exhibiting curve-like shapes resembling formant trajectories, whereas last-layer attention maps produce coarser, blob-like high-value regions.

\subsection{Evaluating GradCAM and LayerCAM using Modified RISE-eval: Masking performance analysis}~\label{sec:evaluation_analysis}

\begin{figure*}[htbp] % 允许 Here、Top、Bottom、Page 放置
    \centering
    \caption{Masking performance curves produced by Modified RISE-eval for GradCAM and LayerCAM, illustrating how the classification accuracy of our speaker recognition network changes as model inputs are progressively masked using attention maps derived from activations at different layers of ResNet34.}  
    \label{fig:performance_curves}
    
    \begin{subfigure}[t]{0.45\textwidth} 
        \centering
        \includegraphics[width=\textwidth, height=180pt]{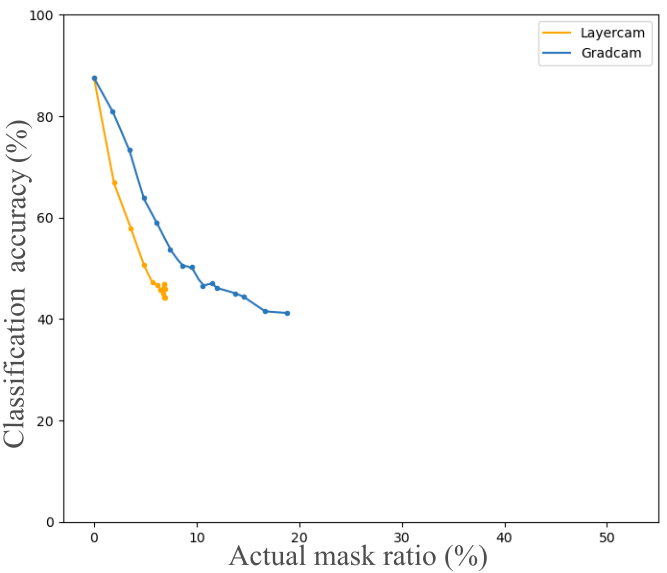}
        \caption{ResNet34 Layer 1}  
        \label{fig:performance_curve_a}
    \end{subfigure}  
    \begin{subfigure}[t]{0.45\textwidth} 
        \centering
        \includegraphics[width=\textwidth, height=180pt]{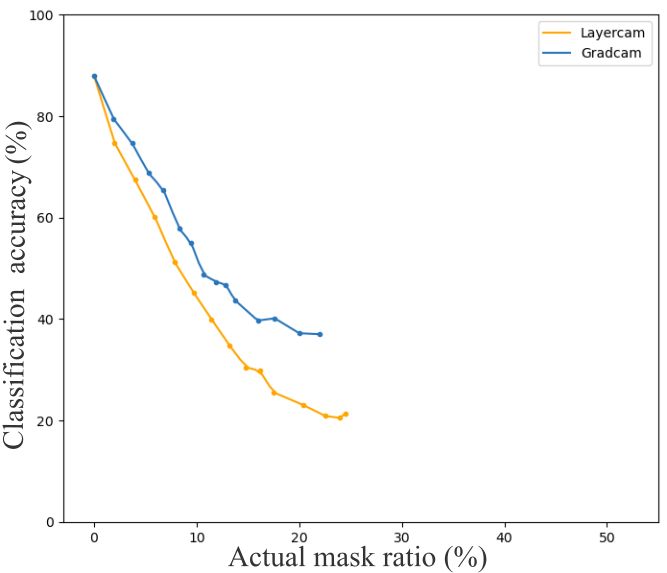}
        \caption{ResNet34 Layer 2}  
        \label{fig:performance_curve_b}
    \end{subfigure}  
    \begin{subfigure}[t]{0.45\textwidth} 
        \centering
        \includegraphics[width=\textwidth, height=180pt]{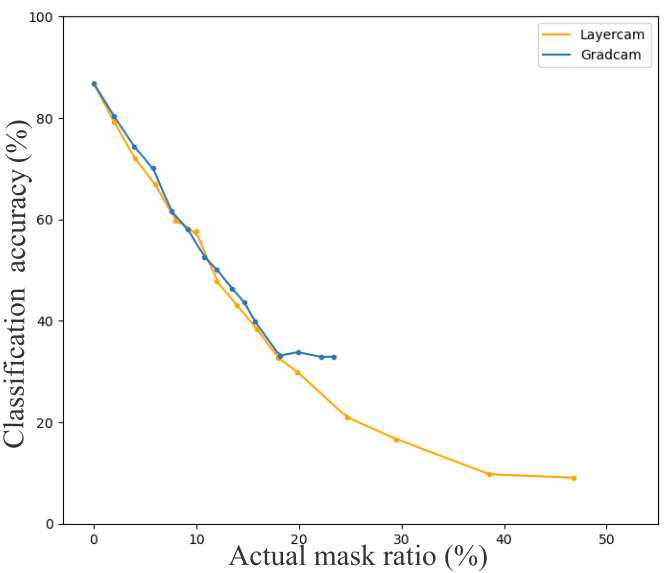}
        \caption{ResNet34 Layer 3}  
        \label{fig:performance_curve_c}
    \end{subfigure}  
    \begin{subfigure}[t]{0.45\textwidth} 
        \centering
        \includegraphics[width=\textwidth, height=180pt]{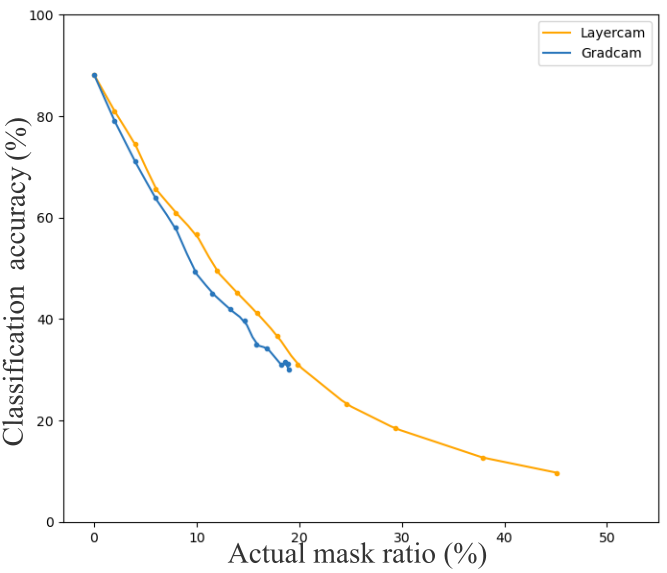}
        \caption{ResNet34 Layer 4}  
        \label{fig:performance_curve_d}
    \end{subfigure}  
\end{figure*}

Nowadays, some studies use the class-discriminative ability of attention maps as a standard for assessing their quality~\cite{selvaraju2016grad, selvaraju2017grad, jiang2021layercam}. This requires manually annotating the regions of specific semantic classes relevant to the network's decision, and then evaluating the consistency between attention maps and these human annotations. In contrast, our proposed Modified RISE-eval algorithm allows the network itself to assess whether the attention map explaining its attention mechanism is correct. Given that the regions highlighted by the attention map are claimed to be the regions that influence, or are important to, the network's decision, Modified RISE-eval directly verifies this claim by progressively masking these regions within the input and quantifying the resulting effect on the network's decision (i.e. change in model performance). The curve depicting how model performance changes as the masking ratio of the input increases is referred to as the masking performance curve. This section comparatively analyse the masking performance curves generated by Modified RISE-eval for the attention maps of GradCAM and LayerCAM, which reveal the attention mechanism of our speaker recognition network, as shown in Fig.~\ref{fig:performance_curves}.

% if the highlighted regions (i.e. given by attention maps) truly reflect what the network considers important for its decisions, masking these regions from inputs should cause a rapid decline in performance. 

% Both qualitative and quantitative attention map analysis provides a deeper understanding of the visualization methods. In this subsection, we further evaluate all attention maps we visualized during the quantitative analysis process using the proposed Modified RISE-eval method. In more detail, the Modified RISE-eval can generate intermediate evaluation results, the masking performance curves, and final evaluation scores for different visualization methods of different setups. 

% Before generating the final evaluation score for each visualization method, both RISE-eval of our version and that of P et al. generate intermediate evaluation results: the masking performance curve. This curve helps to observe the quality of the visualization method intuitively. Specifically, assuming that a better method causes model performance to change more rapidly, we can compare the qualities of different visualization methods by observing the positions of different curves. Our version avoids over-masking the model input when the attention map lacks sufficient important pixels. This modification results in a different curve to be observed, also requiring a different approach to calculating the evaluation score. This section will introduce the masking performance curve generated when using our RISE-eval to evaluate GradCAM and LayerCAM on the contrastive-learned speaker recognition neural network, and how the corresponding evaluation scores are derived.

There are four subfigures inside Fig.~\ref{fig:performance_curves}. The x-axis of each subfigure shows the average mask degree applied to the model inputs (i.e. $R^{\mathrm{mask}}$), while the y-axis shows the classification accuracy of our speaker recognition network in separately classifying inputs masked to each averaged mask ratio in $R^{\mathrm{mask}}$ (i.e. $P$). In Fig.~\ref{fig:performance_curve_a}, the blue and yellow curves are the masking performance curves of GradCAM and LayerCAM, respectively, with attention maps derived from activations of the first ResNet34 layer. Figs.~\ref{fig:performance_curve_b}, \ref{fig:performance_curve_c}, and \ref{fig:performance_curve_d} follow the same colour convention, with attention maps derived from activations of the second, third, and fourth ResNet34 layers, respectively.

All curves across the four subplots exhibit a clear effective interval: as the masking ratio applied to the model input increases, classification accuracy declines consistently, and once the ratio rises from 0 to a certain averaged mask degree, this decline flattens and eventually stabilises. This behaviour arises directly from our design: Modified RISE-eval restricts sampling to pixels in the attention map whose values exceed the threshold $t$ ($t=0.2$) for input masking, rather than sampling all pixels of the attention map, including pixels with low or negligible values, to force the masking ratio to 100\%, as RISE-eval~\cite{petsiuk2018rise, li2022reliable} does (see curves in Fig.~\ref{fig:li_result}).

Fig.~\ref{fig:performance_curve_a} shows the masking performance curves obtained from the first-layer attention maps of GradCAM and LayerCAM, respectively. It is evident that, within the overlapping portion of their effective intervals, the masking performance curve of LayerCAM lies consistently below that of GradCAM, indicating that masking the input based on LayerCAM's attention maps causes a substantially greater degradation in model performance than masking based on GradCAM. This suggests that, when based on first-layer ResNet activations and evaluated over the same number of pixels, LayerCAM outperforms GradCAM in analysing and visualising network attention. In addition, the effective interval of the GradCAM curve is longer, indicating that a greater number of pixels in its attention map exceed the predefined threshold.

Fig.~\ref{fig:performance_curve_b} shows the masking performance curves obtained as the input is progressively masked based on the second-layer attention maps of GradCAM and LayerCAM, respectively. Consistent with the first-layer results, within the overlapping portion of their effective intervals, the masking performance curve of LayerCAM lies consistently below that of GradCAM, indicating that LayerCAM's quality advantage also holds for second-layer activations. Unlike the first-layer case, however, the effective interval of the LayerCAM curve is longer at the second layer, indicating that, compared with GradCAM, LayerCAM's second-layer attention maps contain a greater number of pixels exceeding the predefined threshold.

Fig.~\ref{fig:performance_curve_c} shows the masking performance curves obtained as the input is progressively masked based on the third-layer attention maps of GradCAM and LayerCAM, respectively. Unlike the results from Fig.~\ref{fig:performance_curve_a} and \ref{fig:performance_curve_b}, within the overlapping portion of their effective intervals, the masking performance curves of LayerCAM and GradCAM are largely coincident. This overlap suggests that, when based on third-layer ResNet activations and evaluated over the same number of pixels, the two methods are broadly comparable in their ability to analyse and visualise network attention. In addition, LayerCAM's third-layer attention maps still contain a greater number of pixels exceeding the predefined  threshold.

Fig.~\ref{fig:performance_curve_d} shows the masking performance curves obtained as the input is progressively masked based on the fourth-layer attention maps of GradCAM and LayerCAM, respectively. Unlike the results from the first three layers, within the overlapping portion of their effective intervals, the masking performance curve of GradCAM instead lies below that of LayerCAM, indicating that masking based on GradCAM's attention maps causes a greater degradation in model performance than masking based on LayerCAM. This suggests that, when based on fourth-layer ResNet activations and evaluated over the same number of pixels, GradCAM outperforms LayerCAM in analysing and visualising network attention. In addition, LayerCAM's fourth-layer attention maps contain a greater number of pixels exceeding the predefined threshold.

In summary, to verify whether the regions selected by attention maps genuinely influence the decisions of our speaker recognition network, input masking is performed using the proposed Modified RISE-eval. According to the results, LayerCAM outperforms GradCAM at the activations of the first two layers (i.e. the first and second ResNet layers), causing a greater degradation in model performance when used for masking. At the third layer, the two methods perform comparably, while at the fourth layer GradCAM overtakes LayerCAM, suggesting that the influence of LayerCAM's attention maps on model decisions diminishes progressively as the ResNet layer depth increases. It should be noted that all of the above conclusions are drawn when the same number of pixels is considered. In terms of the number of pixels involved in the input masking process, LayerCAM consistently retains more pixels exceeding the predefined threshold from the second layer onwards. It is worth noting, however, that a larger number of valid pixels does not necessarily correlate with better attention map quality.

\subsection{Evaluating GradCAM and LayerCAM using Modified RISE-eval: Evaluation scores}~\label{sec:evaluation_scores}

\begin{table}[h!]
\centering
\caption{Final evaluation scores of LayerCAM and GradCAM produced by Modified RISE-eval, where Layer 1–4 refer to the four ResNet layers where activations are used to visualise the attention maps.}
\begin{tabular}{lcccc}
\toprule
 & \textbf{Layer 1} & \textbf{Layer 2} & \textbf{Layer 3} & \textbf{Layer 4} \\
\midrule
\textbf{LayerCAM} & 15.50 & 12.58 & 10.27 & 9.52 \\
\textbf{GradCAM} & 10.14 & 9.40 & 8.90 & 10.12 \\
\bottomrule
\end{tabular}
\label{tab:layer_comparsion}
\end{table}

% \begin{tabular}{lccccc}
% \toprule
%  & \textbf{Layer 1} & \textbf{Layer 2} & \textbf{Layer 3} & \textbf{Layer 4} & \textbf{Overall} \\
% \midrule
% \textbf{LayerCAM} & \textbf{15.50} & 12.58 & 10.27 & 9.52 & \textbf{11.97} \\
% \textbf{GradCAM} & \text{10.14} & 9.40 & 8.90 & \textbf{10.12} & 9.64 \\
% \bottomrule
% \end{tabular}
% \label{tab:layer_comparsion}
% \end{table}

In Section~\ref{sec:evaluation_analysis}, evaluating the quality of attention maps and their corresponding visualisation methods requires comparing the complete masking performance curve for each case individually. Modified RISE-eval offers the functionality to score each masking performance curve, quantifying the attention map quality into a single metric for quick comparison. This section discusses the scores assigned by Modified RISE-eval to the eight attention maps corresponding to the eight masking performance curves shown in Fig.~\ref{fig:performance_curves}, namely the GradCAM and LayerCAM attention maps derived from the activations of four ResNet layers, with the results presented in Table~\ref{tab:layer_comparsion}.

In Table~\ref{tab:layer_comparsion}, the scores assigned by Modified RISE-eval to the GradCAM and LayerCAM attention maps are presented. Consistent with the masking performance curves in Fig.~\ref{fig:performance_curves}, LayerCAM outperforms GradCAM across the first three ResNet layers, whilst GradCAM overtakes LayerCAM at the fourth layer. In conclusion, our experimental results on the speaker recognition task reveal that GradCAM is recommended over LayerCAM when using activations from the final layer, which is closest to the model's decision, as its attention maps better highlight regions that genuinely influence model decisions (i.e. as verified through input masking). For activations from the majority of layers (i.e. the first three ResNet layers), LayerCAM is the more suitable choice for analysing the attention mechanism of the speaker recognition network.
% 值得注意的是，以上结果和图片分类领域不成文的一致，他们默认使用GradCAM发现LayerCAM的优势在于在浅层激活上进行使用，GradCAM的优势在于靠近决策层的激活上进行使用。

\section{Discussion}

Rather than extending our experiments to more tasks and neural network architectures, two academic questions warrant deeper investigation. The first concerns the ultimate purpose of attention map visualisation. Numerous visualisation methods currently exist, each approaching the problem from a distinct perspective (i.e. sensitivity, importance, or selectivity) and sometimes from a combination of multiple perspectives (e.g. CAM-based methods). However, even with evaluation methods that help identify which perspective is better or more faithful, we still lack a principled paradigm for interpreting  what the network's attention mechanism truly represents with reference to the visualised attention maps. More importantly, it remains unclear what tangible benefits interpreting a network's attention mechanism can offer to humans in practice. This renders the topic of studying network's attention theoretically valuable yet of limited practical utility.

Second, as shown in Table~\ref{tab:layer_comparsion}, both GradCAM and LayerCAM achieve their highest evaluation scores at the first ResNet layer. However, taken together with the results of Section~\ref{sec:qualitative_analysis}, last-layer attention maps (i.e. particularly those of GradCAM) exhibit stronger class-discriminative ability, revealing a discrepancy between input-masking-based evaluation scores and class-discriminative ability. We attribute this to the following: deeper-layer attention maps tend to be coarser and cover larger regions (as discussed in Section~\ref{sec:qualitative_analysis}), making them more likely to highlight silent regions of the spectrogram input; shallower-layer attention maps, by contrast, are finer-grained and more closely resemble the input features, highlighting non-silent regions such as formants. Consequently, at the same masking ratio, shallower-layer attention maps mask more formant-rich regions than deeper-layer attention maps, introducing an inherent bias in Modified RISE-eval towards shallower-layer attention maps. Addressing this bias problem of Modified RISE-eval remains a direction for future work.

\section{Conclusion}
This work proposes to analyse and visualise the attention mechanism of neural networks. Specifically, GradCAM and LayerCAM are applied to analyse and visualise what input regions are selectively processed during the decision-making process of a ResNet34-based speaker recognition network, with representative cases presented to demonstrate the attention maps produced by both methods. Notably, the results show that GradCAM, when using activations from the final ResNet layer, exhibits strong class-discriminative ability, effectively localising input regions relevant to target speaker.

Furthermore, this work revisits the underexplored RISE-eval algorithm and proposes Modified RISE-eval to address two limitations of the original: indiscriminative evaluation results and the introduction of irrelevant evaluation factors. Modified RISE-eval is applied to evaluate the quality of GradCAM and LayerCAM attention maps, producing masking performance curves with effective intervals and corresponding evaluation scores. The results indicate that, for the final ResNet layer activations, GradCAM attention maps better highlight input regions that genuinely influence model decisions, causing a greater degradation in model performance during input masking than LayerCAM; while for the remaining layers, LayerCAM better analyses and visualises the network's attention by highlighting input regions that more effectively affect model decisions during input masking.

\section*{Acknowledgment}
I would like to thank Chung et al. and Nagrani et al. for releasing the VoxCeleb dataset and publishing their pretrained neural network. I also appreciate the work of Li et al., which has greatly contributed to the discussion in this work. I am grateful to Professor Philip Jackson and Jon Barker for their valuable suggestions in strengthening this work. Finally, a qualitative analysis criterion for the class-discriminative ability of attention maps can be found in the PhD thesis of the first author.

\ifCLASSOPTIONcaptionsoff
  \newpage
\fi



\bibliographystyle{IEEEtran}
% \bibliography{IEEEabrv,../bib/paper}
\bibliography{bibtex/bib/IEEEexample}
\end{document}